\documentclass[twocolumn,aps,tightenlinefs,superscriptaddress,preprintnumbers]{revtex4}
\usepackage{graphicx,color,dcolumn,booktabs,bm}
\usepackage{longtable,lscape}
\usepackage{amsmath}
\usepackage{indentfirst}
\usepackage{epsfig}
\usepackage{feynmf}   
\usepackage{epstopdf}   
\usepackage{slashed}  
\usepackage{cases}
\usepackage{color}
\usepackage{multirow}
\usepackage{ulem}
\usepackage{graphicx,color,dcolumn,booktabs,bm}
\usepackage{verbatim}
\usepackage[colorlinks,linkcolor=black,anchorcolor=green,citecolor=blue]{hyperref}
\usepackage{mathrsfs}
\usepackage{rotating}
\usepackage{threeparttable}
\usepackage{dcolumn}

\newcommand{\lum}{{\cal L}}

\newcommand{\BR}{{\cal B}}

\newcommand{\pip}{\pi^+}
\newcommand{\pim}{\pi^-}

\newcommand{\EE}{e^+e^-}

\newcommand{\beq}{\begin{equation}}
\newcommand{\eeq}{\end{equation}}
\newcommand{\bitm}{\begin{itemize}}
\newcommand{\eitm}{\end{itemize}}

\newcommand{\R}{R^{++}}
\newcommand{\ones}{\Upsilon(1S)}
\newcommand{\twos}{\Upsilon(2S)}
\newcommand{\onetwos}{\Upsilon(1S,2S)}

\newcommand{\fours}{\Upsilon(4S)}
\newcommand{\fives}{\Upsilon(5S)}

\newcolumntype{d}[1]{D{.}{.}{#1}}

\begin{document}
\hyphenpenalty=10000


\title{\quad\\[0.1cm]\boldmath Search for a doubly-charged $DDK$ bound state in
	$\onetwos$ inclusive decays and via direct production in $e^+e^-$ collisions
	at $\sqrt{s}$ = 10.520, 10.580, and 10.867~GeV}

\noaffiliation
\affiliation{University of the Basque Country UPV/EHU, 48080 Bilbao}
\affiliation{Beihang University, Beijing 100191}
\affiliation{University of Bonn, 53115 Bonn}
\affiliation{Brookhaven National Laboratory, Upton, New York 11973}
\affiliation{Budker Institute of Nuclear Physics SB RAS, Novosibirsk 630090}
\affiliation{Faculty of Mathematics and Physics, Charles University, 121 16 Prague}
\affiliation{Chonnam National University, Gwangju 61186}
\affiliation{University of Cincinnati, Cincinnati, Ohio 45221}
\affiliation{Deutsches Elektronen--Synchrotron, 22607 Hamburg}
\affiliation{University of Florida, Gainesville, Florida 32611}
\affiliation{Department of Physics, Fu Jen Catholic University, Taipei 24205}
\affiliation{Key Laboratory of Nuclear Physics and Ion-beam Application (MOE) and Institute of Modern Physics, Fudan University, Shanghai 200443}
\affiliation{Justus-Liebig-Universit\"at Gie\ss{}en, 35392 Gie\ss{}en}
\affiliation{Gifu University, Gifu 501-1193}
\affiliation{SOKENDAI (The Graduate University for Advanced Studies), Hayama 240-0193}
\affiliation{Gyeongsang National University, Jinju 52828}
\affiliation{Department of Physics and Institute of Natural Sciences, Hanyang University, Seoul 04763}
\affiliation{University of Hawaii, Honolulu, Hawaii 96822}
\affiliation{High Energy Accelerator Research Organization (KEK), Tsukuba 305-0801}
\affiliation{J-PARC Branch, KEK Theory Center, High Energy Accelerator Research Organization (KEK), Tsukuba 305-0801}
\affiliation{Higher School of Economics (HSE), Moscow 101000}
\affiliation{Forschungszentrum J\"{u}lich, 52425 J\"{u}lich}
\affiliation{IKERBASQUE, Basque Foundation for Science, 48013 Bilbao}
\affiliation{Indian Institute of Science Education and Research Mohali, SAS Nagar, 140306}
\affiliation{Indian Institute of Technology Bhubaneswar, Satya Nagar 751007}
\affiliation{Indian Institute of Technology Guwahati, Assam 781039}
\affiliation{Indian Institute of Technology Hyderabad, Telangana 502285}
\affiliation{Indian Institute of Technology Madras, Chennai 600036}
\affiliation{Indiana University, Bloomington, Indiana 47408}
\affiliation{Institute of High Energy Physics, Chinese Academy of Sciences, Beijing 100049}
\affiliation{Institute of High Energy Physics, Vienna 1050}
\affiliation{Institute for High Energy Physics, Protvino 142281}
\affiliation{INFN - Sezione di Napoli, 80126 Napoli}
\affiliation{INFN - Sezione di Torino, 10125 Torino}
\affiliation{Advanced Science Research Center, Japan Atomic Energy Agency, Naka 319-1195}
\affiliation{J. Stefan Institute, 1000 Ljubljana}
\affiliation{Institut f\"ur Experimentelle Teilchenphysik, Karlsruher Institut f\"ur Technologie, 76131 Karlsruhe}
\affiliation{Kavli Institute for the Physics and Mathematics of the Universe (WPI), University of Tokyo, Kashiwa 277-8583}
\affiliation{Kennesaw State University, Kennesaw, Georgia 30144}
\affiliation{Department of Physics, Faculty of Science, King Abdulaziz University, Jeddah 21589}
\affiliation{Kitasato University, Sagamihara 252-0373}
\affiliation{Korea Institute of Science and Technology Information, Daejeon 34141}
\affiliation{Korea University, Seoul 02841}
\affiliation{Kyoto Sangyo University, Kyoto 603-8555}
\affiliation{Kyungpook National University, Daegu 41566}
\affiliation{Universit\'{e} Paris-Saclay, CNRS/IN2P3, IJCLab, 91405 Orsay}
\affiliation{P.N. Lebedev Physical Institute of the Russian Academy of Sciences, Moscow 119991}
\affiliation{Liaoning Normal University, Dalian 116029}
\affiliation{Faculty of Mathematics and Physics, University of Ljubljana, 1000 Ljubljana}
\affiliation{Ludwig Maximilians University, 80539 Munich}
\affiliation{Luther College, Decorah, Iowa 52101}
\affiliation{Malaviya National Institute of Technology Jaipur, Jaipur 302017}
\affiliation{University of Maribor, 2000 Maribor}
\affiliation{Max-Planck-Institut f\"ur Physik, 80805 M\"unchen}
\affiliation{School of Physics, University of Melbourne, Victoria 3010}
\affiliation{University of Mississippi, University, Mississippi 38677}
\affiliation{University of Miyazaki, Miyazaki 889-2192}
\affiliation{Moscow Physical Engineering Institute, Moscow 115409}
\affiliation{Graduate School of Science, Nagoya University, Nagoya 464-8602}
\affiliation{Universit\`{a} di Napoli Federico II, 80126 Napoli}
\affiliation{Nara Women's University, Nara 630-8506}
\affiliation{National Central University, Chung-li 32054}
\affiliation{National United University, Miao Li 36003}
\affiliation{Department of Physics, National Taiwan University, Taipei 10617}
\affiliation{H. Niewodniczanski Institute of Nuclear Physics, Krakow 31-342}
\affiliation{Nippon Dental University, Niigata 951-8580}
\affiliation{Niigata University, Niigata 950-2181}
\affiliation{Novosibirsk State University, Novosibirsk 630090}
\affiliation{Osaka City University, Osaka 558-8585}
\affiliation{Pacific Northwest National Laboratory, Richland, Washington 99352}
\affiliation{Panjab University, Chandigarh 160014}
\affiliation{Peking University, Beijing 100871}
\affiliation{University of Pittsburgh, Pittsburgh, Pennsylvania 15260}
\affiliation{Punjab Agricultural University, Ludhiana 141004}
\affiliation{Research Center for Nuclear Physics, Osaka University, Osaka 567-0047}
\affiliation{Meson Science Laboratory, Cluster for Pioneering Research, RIKEN, Saitama 351-0198}
\affiliation{Department of Modern Physics and State Key Laboratory of Particle Detection and Electronics, University of Science and Technology of China, Hefei 230026}
\affiliation{Seoul National University, Seoul 08826}
\affiliation{Showa Pharmaceutical University, Tokyo 194-8543}
\affiliation{Soochow University, Suzhou 215006}
\affiliation{Soongsil University, Seoul 06978}
\affiliation{Sungkyunkwan University, Suwon 16419}
\affiliation{School of Physics, University of Sydney, New South Wales 2006}
\affiliation{Department of Physics, Faculty of Science, University of Tabuk, Tabuk 71451}
\affiliation{Tata Institute of Fundamental Research, Mumbai 400005}
\affiliation{Department of Physics, Technische Universit\"at M\"unchen, 85748 Garching}
\affiliation{School of Physics and Astronomy, Tel Aviv University, Tel Aviv 69978}
\affiliation{Department of Physics, Tohoku University, Sendai 980-8578}
\affiliation{Earthquake Research Institute, University of Tokyo, Tokyo 113-0032}
\affiliation{Department of Physics, University of Tokyo, Tokyo 113-0033}
\affiliation{Tokyo Institute of Technology, Tokyo 152-8550}
\affiliation{Tokyo Metropolitan University, Tokyo 192-0397}
\affiliation{Utkal University, Bhubaneswar 751004}
\affiliation{Virginia Polytechnic Institute and State University, Blacksburg, Virginia 24061}
\affiliation{Wayne State University, Detroit, Michigan 48202}
\affiliation{Yamagata University, Yamagata 990-8560}
\affiliation{Yonsei University, Seoul 03722}
\author{Y.~Li}\affiliation{Beihang University, Beijing 100191} 
\author{S.~Jia}\affiliation{Key Laboratory of Nuclear Physics and Ion-beam Application (MOE) and Institute of Modern Physics, Fudan University, Shanghai 200443} 
\author{C.~P.~Shen}\affiliation{Key Laboratory of Nuclear Physics and Ion-beam Application (MOE) and Institute of Modern Physics, Fudan University, Shanghai 200443} 
\author{I.~Adachi}\affiliation{High Energy Accelerator Research Organization (KEK), Tsukuba 305-0801}\affiliation{SOKENDAI (The Graduate University for Advanced Studies), Hayama 240-0193} 
\author{H.~Aihara}\affiliation{Department of Physics, University of Tokyo, Tokyo 113-0033} 
\author{S.~Al~Said}\affiliation{Department of Physics, Faculty of Science, University of Tabuk, Tabuk 71451}\affiliation{Department of Physics, Faculty of Science, King Abdulaziz University, Jeddah 21589} 
\author{D.~M.~Asner}\affiliation{Brookhaven National Laboratory, Upton, New York 11973} 
\author{T.~Aushev}\affiliation{Higher School of Economics (HSE), Moscow 101000} 
\author{R.~Ayad}\affiliation{Department of Physics, Faculty of Science, University of Tabuk, Tabuk 71451} 
\author{V.~Babu}\affiliation{Deutsches Elektronen--Synchrotron, 22607 Hamburg} 
% ...
\collaboration{The Belle Collaboration}

\begin{abstract}
We report the results of a first search for a doubly-charged $DDK$ bound state, denoted the $R^{++}$, in $\Upsilon(1S)$
and $\Upsilon(2S)$ inclusive decays and via direct production in $e^+e^-$ collisions at $\sqrt{s}$ = 10.520, 10.580, and 10.867~GeV.
The search uses data accumulated with the Belle detector at the KEKB asymmetric-energy $e^+e^-$ collider.
No significant signals are observed in the $D^{+}D_{s}^{*+}$ invariant-mass spectra
of all studied modes. The 90\% credibility level upper limits on their product
branching fractions in $\Upsilon(1S)$ and $\Upsilon(2S)$ inclusive decays (${\cal B}(\Upsilon(1S,2S) \to R^{++} + anything)
\times {\cal B}(R^{++} \to D^{+} D_{s}^{*+})$), and the product values of Born cross section
and branching fraction in $e^+e^-$ collisions ($\sigma(e^+e^- \to R^{++} + anything) \times {\cal B}(R^{++} \to D^{+} D_{s}^{*+})$)
at $\sqrt{s}$ = 10.520, 10.580, and 10.867~GeV under different assumptions of $R^{++}$ masses varying from 4.13 to 4.17 GeV/$c^2$,
and widths varying from 0 to 5~MeV are obtained.
\end{abstract}

\maketitle
\section{\boldmath Introduction}
In 2003, a narrow resonance near 2.32~GeV/$c^{2}$, the $D_{s0}^{*}(2317)^{+}$, was observed
by BaBar~\cite{2317} via its decay to $D_{s}^{+}\pi^0$. The $D_{s0}^{*}(2317)^{+}$ was subsequently confirmed by CLEO~\cite{CLEO2317}
and Belle~\cite{Belle2317}. The observed low mass and narrow width of the $D_{s0}^{*}(2317)^{+}$ strongly disfavor the
interpretation of this state as a $P$-wave $c\bar{s}$ state, both in potential model~\cite{quark1,quark2,quark3,quark4,quark5,quark6}
and lattice Quantum Chromodynamics (QCD)~\cite{lattice1,lattice2} descriptions.
Inclusion of charge-conjugate decays is implicitly assumed throughout this analysis.
Instead, it has been proposed as a possible candidate for a $DK$  molecule~\cite{molecule1,molecule2,molecule3,molecule4,molecule5,molecule6}, a ($cq$)($\bar{s}\bar{q}$)
tetraquark state~\cite{tetra1,tetra2,tetra3}, or a mixture of a $c\bar{s}$ state and
tetraquark~\cite{mixture1,mixture2,mixture3,mixture4,mixture5,mixture6,mixture7,mixture8}.
The absolute branching fraction of $D_{s0}^{*}(2317)^{+}\to D_{s}^{+}\pi^0$
was measured by BESIII to be $1.00^{+0.00}_{-0.14} \pm 0.14$~\cite{ab-br}.
This result indicates that the $D_{s0}^{*}(2317)^{+}$  has a much smaller branching fraction
to $D_{s}^{*+}\gamma$ than to $D_{s}^{+}\pi^0$, and this agrees with the expectation of the
conventional $c\bar{s}$ state hypothesis~\cite{hy1} or the hadronic molecule picture of $DK$~\cite{hy2,hy3}.
There have been theoretical interpretations of the $D_{s0}^{*}(2317)^{+}$ and $D_{s1}(2460)^{+}$ as chiral partners, with production mechanisms related to the spontaneous breaking of chiral symmetry~\cite{chiral1,chiral2}.

By exchanging a kaon, a $D^{+}D_{s0}^{*}(2317)^{+}$ molecular state can be
formed with a binding energy of $5-15$ MeV, regardless of whether the $D_{s0}^{*}(2317)^{+}$ is treated
as a $c\bar{s}$ state or a $DK$ molecule~\cite{DDK1}. In Ref.~\cite{DDK2}, the authors
studied the $DDK$ system in a coupled-channel approach, where an isospin $1/2$ state, denoted the $\R$, is formed at 4140
MeV/$c^2$ when the $D_{s0}^{*}(2317)^{+}$ is generated from the $DK$ subsystem.
The $\R$ can be interpreted as a $D^{+}D_{s0}^{*}(2317)^{+}$ molecule-like state with exotic properties:
doubly-charged and doubly-charmed. Hereinafter we also refer to this predicted state as $R^{++}$.

An $\R$ with the properties described above would be able to decay via $R^{++}\to D^+D_{s0}^{*}(2317)^+$,
where $D_{s0}^{*}(2317)^+\to D^+_s\pi^0$ is an isospin-violating process. The alternative processes
are via triangle diagrams into $R^{++}\to D^{+}D_{s}^{*+}$
and $R^{++}\to D_{s}^{+}D^{*+}$~\cite{DDK2,DDK6,DDK4}. The mass of $R^{++}$ is predicted to be in the range of 4.13 to 4.17~GeV/c$^2$~\cite{DDK4}.
The predicted partial decay width of $R^{++}\to D^{+}D_{s}^{*+}$ is much larger than that of $R^{++}\to D_{s}^{+}D^{*+}$;
they are $\Gamma(R^{++} \to D^{+}D_{s}^{*+})$ = $(2.30-2.49)$~MeV and $\Gamma(R^{++} \to D_{s}^{+}D^{*+})$ = $(0.26-0.29)$~MeV~\cite{DDK4}, respectively.

The question whether $QQ\bar q\bar q$ tetraquarks with two heavy quarks $Q$ and two light
antiquarks $\bar q$ are stable or unstable against decay into two $Q\bar q$ mesons has a long history.
It has been largely undecided, mainly due to a lack of experimental information about the
strength of the interaction between two heavy quarks. The discovery of the doubly-charmed
baryon $\Xi_{cc}^{++}$ by LHCb~\cite{Xicc} has provided the crucial experimental input~\cite{bbud,QQqq}.
In Ref.~\cite{QQqq}, the authors predicted the existence of novel narrow doubly-heavy
tetraquark states of the form $QQ\bar q\bar q$ with the method based on the heavy-quark symmetry,
and found that a doubly-charmed tetraquark with a mass of 4156~MeV/c$^2$ and a $J^P$ of $1^+$
decaying into a final state of $D^{+}D_{s}^{*+}$ can be formed. Thus the $D^{+}D_{s}^{*+}$ final state is
a good channel to search for such a tetraquark state.

In this paper, we search for a doubly-charged $DDK$ bound state in the $D^{+}D_{s}^{*+}$ final state
in $\ones$ and $\twos$ inclusive decays, and via direct production in $\EE$ collisions at $\sqrt{s}$ = 10.520, 10.580,
and 10.867~GeV. We report a search for the $\R$ with masses varying from 4.13 to 4.17~GeV/$c^2$ and widths varying from 0 to 5~MeV.

\section{\boldmath The data sample and the belle detector}
This analysis utilizes (5.74 $\pm$ 0.09) fb$^{-1}$ of data collected at the $\ones$ peak
[(102 $\pm$ 3) million $\ones$ events], (24.91 $\pm$ 0.35) fb$^{-1}$ of data collected at
the $\twos$ peak [(158 $\pm$ 4) million $\twos$ events], a data sample of (89.5 $\pm $ 1.3) fb$^{-1}$
collected at $\sqrt{s}$ =  10.520~GeV, a data sample of (711.0 $\pm$ 10.0) fb$^{-1}$
collected at $\sqrt{s}$ = 10.580~GeV [$\fours$ peak], and a data sample of (121.4 $\pm$ 1.7) fb$^{-1}$
collected at $\sqrt{s}$ = 10.867~GeV [$\fives$ peak]. All the data were collected with
the Belle detector~\cite{detector} operating at the KEKB asymmetric-energy $e^+e^-$
collider~\cite{collider}. The Belle detector is described in detail in Ref.~\cite{detector}.
It is a large-solid-angle magnetic spectrometer consisting of a silicon vertex detector,
a 50-layer central drift chamber (CDC), an array of aerogel threshold Cherenkov counters (ACC),
a barrel-like arrangement of time-of-flight scintillation counters (TOF), and an electromagnetic
calorimeter comprising CsI(TI) crystals (ECL) located inside a superconducting solenoid coil that
provides a $1.5~\hbox{T}$ magnetic field. An iron flux return comprising resistive plate chambers (RPCs)
placed outside the coil is instrumented to detect $K^{0}_{L}$ mesons and to identify muons (KLM).

Monte Carlo (MC) signal samples are generated with {\sc EvtGen}~\cite{evtgen} to determine
signal shapes and efficiencies. Initial-state radiation (ISR) is taken into account by assuming that the
cross sections follow a $1/s$ dependence in $\EE \to R^{++} + anything$ reactions, where $s$ is the
center-of-mass energy squared. The mass of $R^{++}$ is chosen from 4.13 to 4.17~GeV/$c^{2}$
in steps of 2.5~MeV/$c^{2}$, with a width varying from 0 to 5~MeV in steps of 1~MeV.
These events are processed by a detector simulation based on {\sc geant3}~\cite{geant}.

Inclusive MC samples of $\Upsilon(1S,2S)$ decays, $\fours \to B^{+}B^{-}/B^{0}\bar{B}^{0}$, $\fives \to B_{s}^{(*)} \bar{B}_{s}^{(*)}$, and $\EE \to q\bar{q}$ $(q=u, d, s, c)$ at $\sqrt{s}$ = 10.520, 10.580, and 10.867~GeV corresponding to four times the integrated luminosity of data are used to study possible peaking backgrounds.

\section{\boldmath Common Event selection criteria}
For well-reconstructed charged tracks, except those from
$K_{S}^{0} \to \pi^{+} \pi^{-}$ decays, the impact parameters perpendicular to
and along the beam direction with respect to the nominal interaction point (IP)
are required to be less than 0.5~cm and 2~cm, respectively, and the transverse momentum in the
laboratory frame is required to be larger than 0.1~GeV/$c$.
For the particle identification (PID) of a well-reconstructed charged track,
information from different detector subsystems, including specific ionization in the CDC,
time measurement in the TOF, and the response of the ACC, is combined to form a likelihood
${\mathcal L}_i$~\cite{pidcode} for particle species $i$, where $i$ =  $\pi$ or $K$.
Tracks with $R_{K}=\mathcal{L}_{K}/(\mathcal{L}_K+\mathcal{L}_\pi)<0.4$ are identified as
pions with an efficiency of 96\%, while 5\% of kaons are misidentified as pions; tracks
with $R_{K}>0.6$ are identified as kaons with an efficiency of 95\%, while 4\% of pions are
misidentified as kaons. Except for tracks from $K^0_S$ decays, all charged tracks
are required to be positively identified by the above procedures.

An ECL cluster is taken as a photon candidate
if it does not match the extrapolation of any charged track. The energy of the photon
is required to be greater than 50~MeV.

The $K_{S}^{0}$ candidates are first reconstructed from pairs of oppositely charged tracks,
which are treated as pions, with a production vertex significantly separated from the average IP,
then selected using an artificial neural network~\cite{neural}
based on two sets of input variables~\cite{input}. The $\phi$ and $\bar{K}^{*}(892)^{0}$
candidates are reconstructed using $K^{+}K^{-}$ and $K^{-}\pip$ decay modes, respectively.
The invariant masses of the $K_{S}^0$ and $\phi$ candidates are required to be within 7 MeV/$c^{2}$ of
the corresponding nominal masses ($>$ 90\% signal events are retained).

We reconstruct $D^{+}$ mesons in the $K^{-} \pip \pip$ and $K_{S}^{0}(\to \pip\pim) \pip$ decay channels,
and $D_{s}^{+}$ mesons in the $\phi\pip$ and $\bar{K}^{*}(892)^{0}K^{+}$ decay channels.
We perform vertex- and mass-constrained fits for $D^{+}$ and $D_{s}^{+}$ candidates, and require
$\chi^{2}_{\rm vertex}/n.d.f.<20$ ($>$ 97\% selection efficiency according to MC simulation).
The selected $D_{s}^{+}$ candidate is combined with a photon to form a $D_{s}^{*+}$ candidate,
and a mass-constrained fit is performed to improve its momentum resolution.

The signal mass windows for $\bar{K}^{*}(892)^{0}$, $D^+$, $D_s^+$, and $D_{s}^{*+}$ candidates have been optimized by maximizing the Punzi parameter $S/(3/2+\sqrt{B})$~\cite{Punzi}. Here, $S$ is the number of $\R$ signal events in the MC-simulated $\twos \to \R +anything $ sample with the mass and width of $\R$ fixed at 4.13 GeV/$c^2$ and 2 MeV assuming $\BR(\twos \to R^{++} + anything) \times \BR(R^{++} \to D^{+} D_{s}^{*+}) = 10^{-4}$, and $B$ is the number of background events in the $\R$ signal window. The number of background events is obtained from the normalized $M_{D^{+}}$ and $M_{D_{s}^{*+}}$ sidebands in the data requiring 4.12 GeV/$c^2$<$M_{D^+D_{s}^{*+}}$<4.14 GeV/$c^2$ as the $\R$ signal region (about 3$\sigma$ according to signal MC simulations).
The optimized signal regions are $|M_{K^{-}\pip} - m_{\bar{K}^{*}(892)^{0}}| < 60$ MeV/$c^{2}$, $|M_{K^{-} \pip \pip/K_{S}^{0}\pip} - m_{D^{+}}| < 6$ MeV/$c^{2}$, $|M_{\phi\pip/\bar{K}^{*}(892)^{0}K^{+}} - m_{D_{s}^{+}}| < 6$ MeV/$c^{2}$, and $|M_{\gamma D_{s}^{+}} - m_{D_{s}^{*+}}| < 9$ MeV/$c^{2}$ for $\bar{K}^{*}(892)^{0}$, $D^+$, $D_s^+$, and $D_{s}^{*+}$ candidates ($>$ 80\% signal events are retained for each intermediate state), respectively, where $m_{\bar{K}^{*}(892)^{0}}$, $m_{{D}_{s}^{+}}$, $m_{D^{+}}$, and $m_{D_{s}^{*+}}$ are the nominal masses of $\bar{K}^{*}(892)^{0}$, ${D}_{s}^{+}$, $D^{+}$, and $D_{s}^{*+}$ mesons~\cite{PDG}. For the process $\ones \to \R + anything$ and $\EE \to R^{++} + anything$ at $\sqrt{s}$ = 10.520, 10.580, and 10.867 GeV, the optimized signal regions of intermediate states are the same.

Finally, when the $D^{+}$ and $D_{s}^{*+}$ candidates are combined to form $R^{++}$ candidates,  all the combinations are preserved for further analysis.
The fraction of events where multiple combinations are selected as $\R$ candidates is 14\% in data, which is consistent with the MC simulation.

\section{\boldmath $\onetwos \to R^{++}$ + anything }
\begin{figure*}[htbp]
	\begin{center}
		\includegraphics[width=5.9cm]{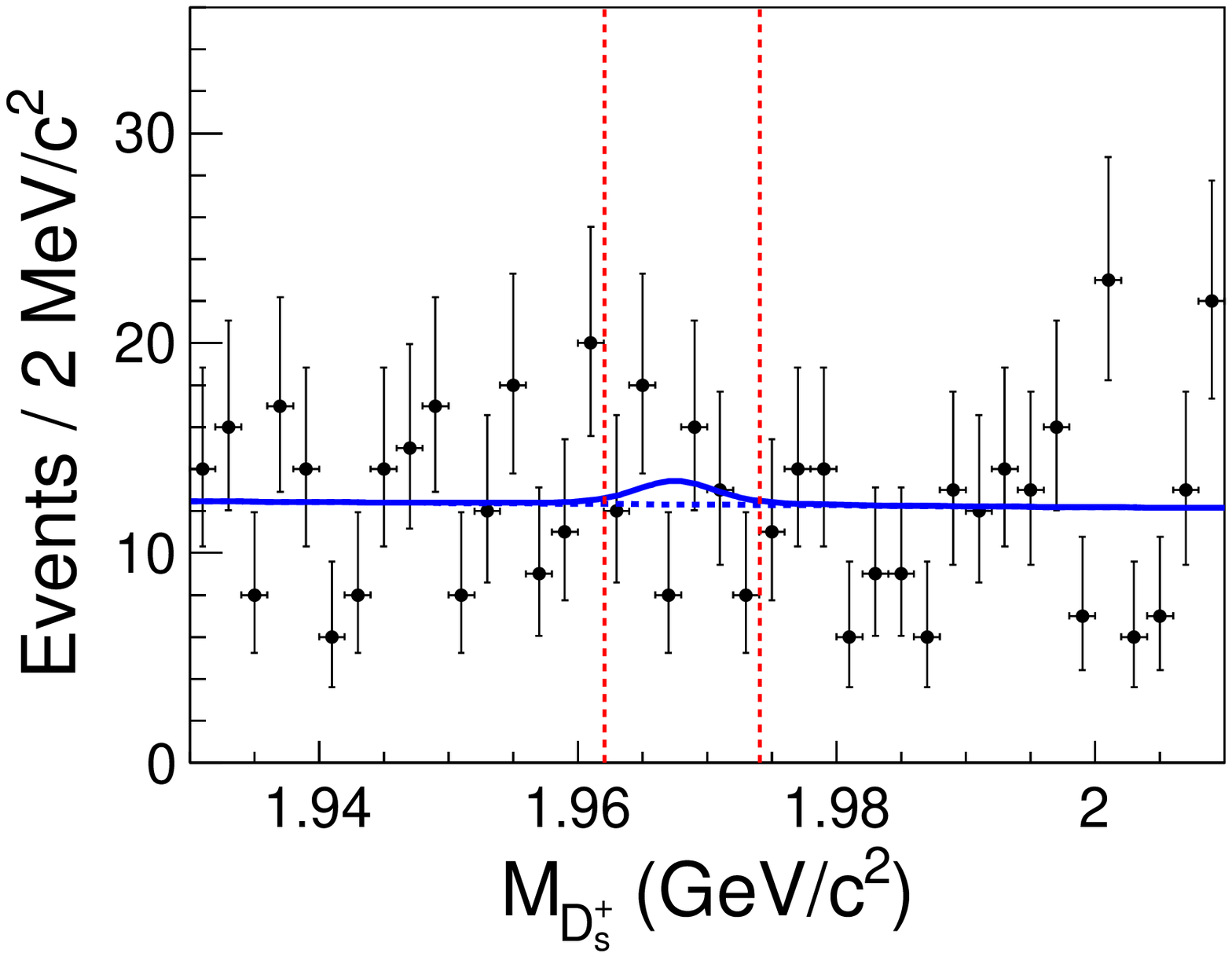}
		\includegraphics[width=5.9cm]{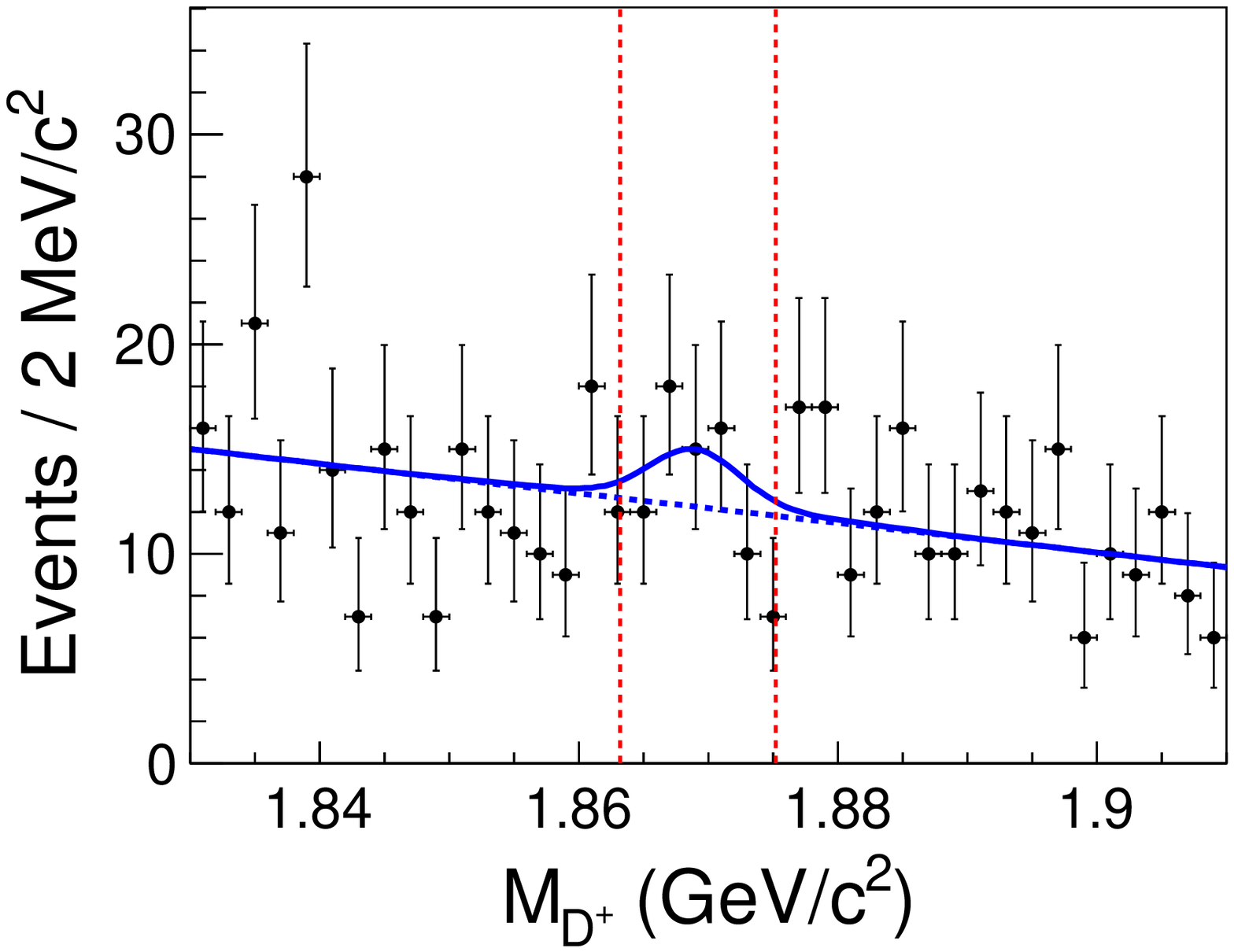}
		\includegraphics[width=5.9cm]{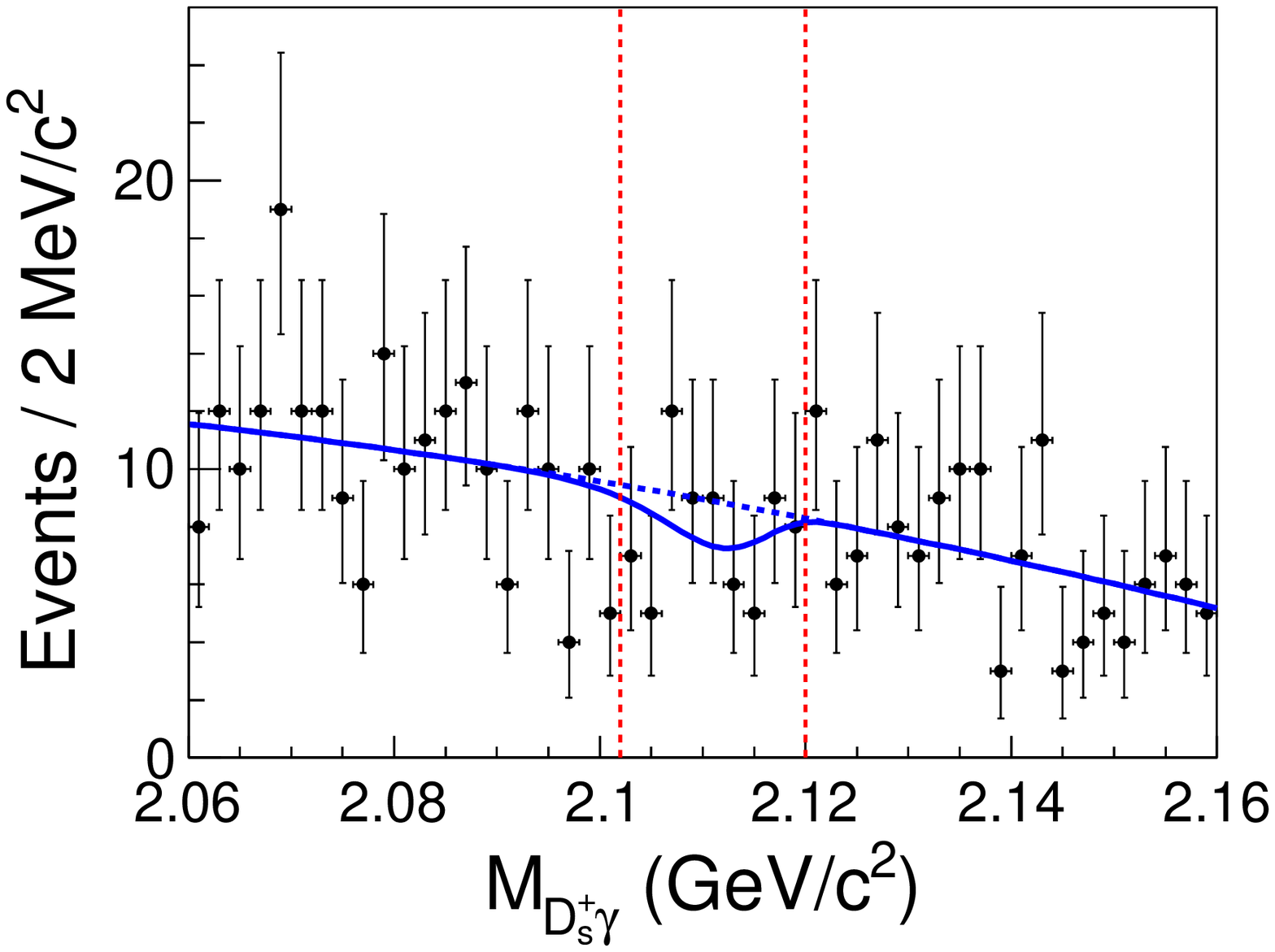}
		\put(-380,102){\bf (a)} \put(-210,102){\bf (b)}  \put(-38,102){\bf (c)}
		\caption{The invariant-mass spectra of the (a) $D_{s}^{+}$, (b) $D^{+}$, and (c) $D_{s}^{*+}$ candidates summed over four reconstructed modes from $\Upsilon(1S)$ data. The points with error bars represent the data, the solid curves show the results of the best fits to the data, and the blue dashed curves are the fitted backgrounds.
			The red dashed lines show the required signal regions.}\label{int_mass_1s}
	\end{center}
\end{figure*}

\begin{figure*}[htbp]
	\begin{center}
		\includegraphics[width=5.9cm]{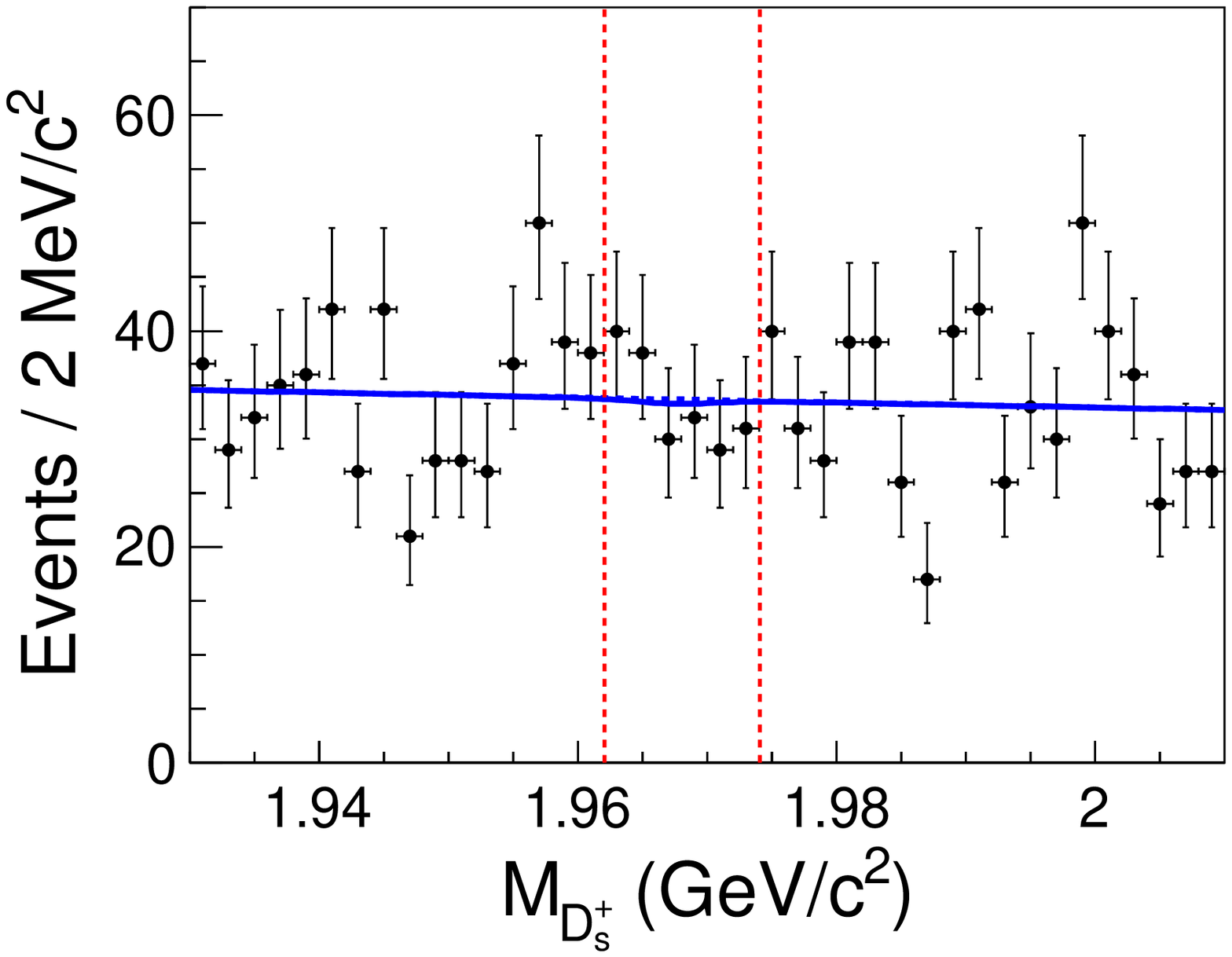}
		\includegraphics[width=5.9cm]{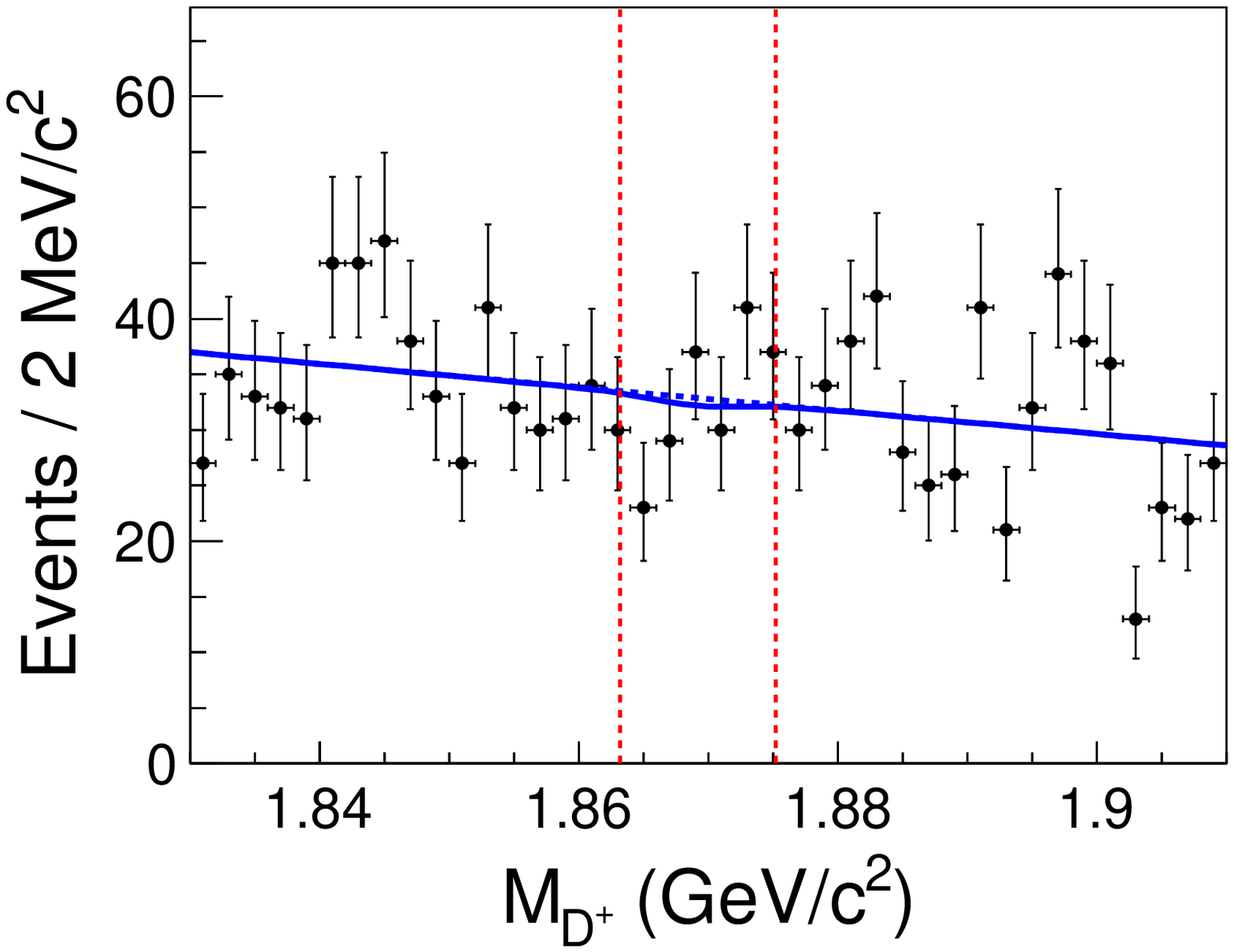}
		\includegraphics[width=5.9cm]{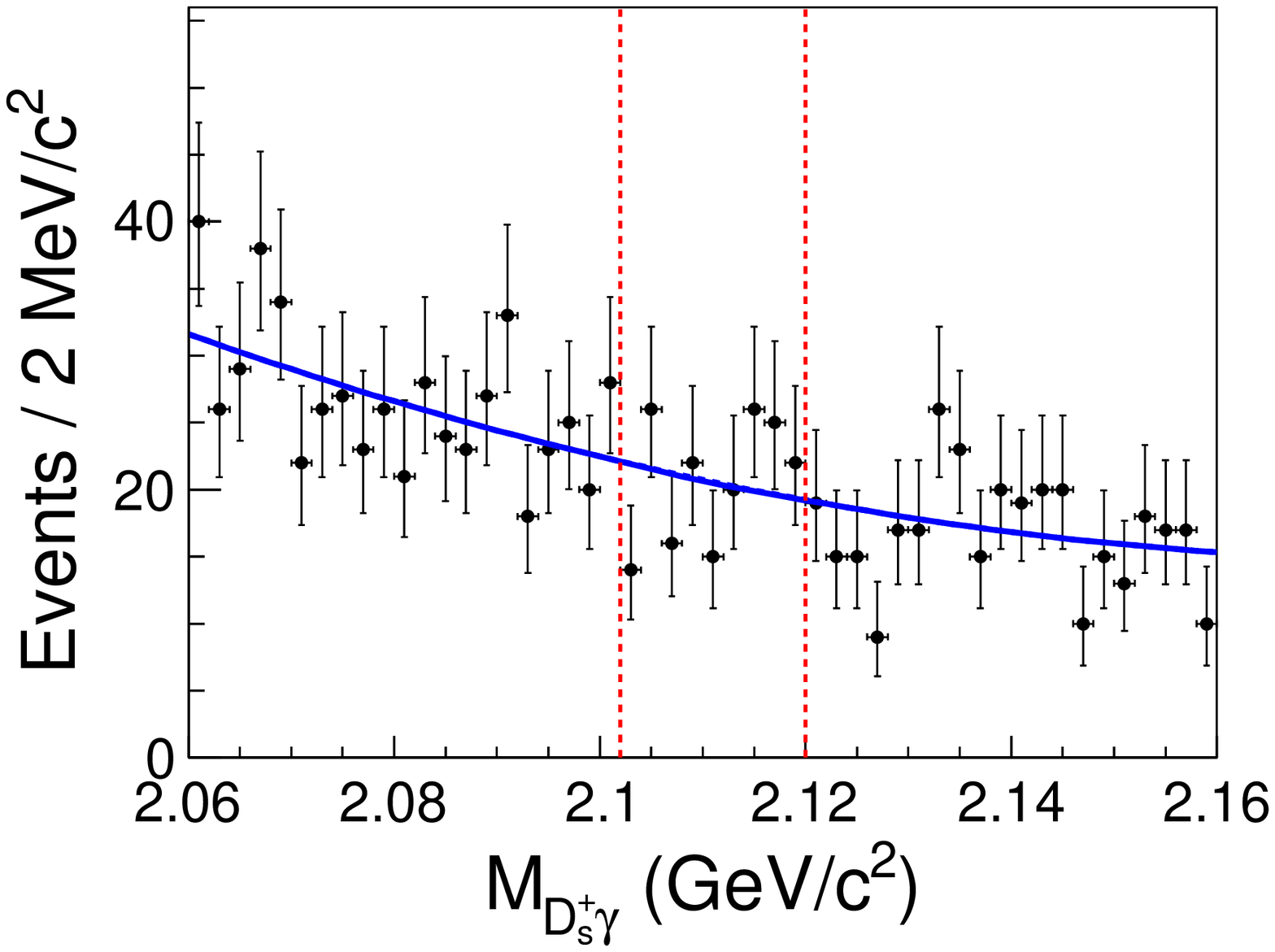}
		\put(-380,102){\bf (a)} \put(-210,102){\bf (b)}  \put(-38,102){\bf (c)}
		\caption{The invariant-mass spectra of the (a) $D_{s}^{+}$, (b) $D^{+}$, and (c) $D_{s}^{*+}$ candidates summed over  four
			reconstructed modes from $\Upsilon(2S)$ data. The points with error bars represent the data,
			the solid curves show the results of the best fits to the data, and the blue dashed curves are the fitted backgrounds.
			The red dashed lines show the required signal regions.
		}\label{int_mass_2s}
	\end{center}
\end{figure*}

In this section, we search for the doubly-charged $DDK$ bound state in $\ones$ and $\twos$ inclusive decays.
After applying the aforementioned common event selections, the invariant-mass distributions of the $D_{s}^{+}$, $D^{+}$,
and $D_{s}^{*+}$ candidates from the $\ones$ and $\twos$ data samples are shown in Figs.~\ref{int_mass_1s} and~\ref{int_mass_2s}, respectively,
together with results of the fits described below.
When drawing each distribution, the signal mass windows of other intermediate states are required.
No clear $D_{s}^{+}$, $D^{+}$, and $D_{s}^{*+}$ signals are observed.
In the fits, the $D_{s}^{+}$ and $D^{+}$ signal shapes are described by double-Gaussian functions, and the $D_{s}^{*+}$ signal
shape is described by a Novosibirsk function~\cite{Novosibirsk}, where the values of parameters are fixed to those obtained
from the fits to the corresponding signal MC distributions. The backgrounds are parametrized by first-order
polynomial functions for $D_{s}^{+}$ and $D^{+}$, and a second-order polynomial function for $D_{s}^{*+}$.

\begin{figure*}[htbp]
	\begin{center}
		\includegraphics[height=6cm,width=8cm]{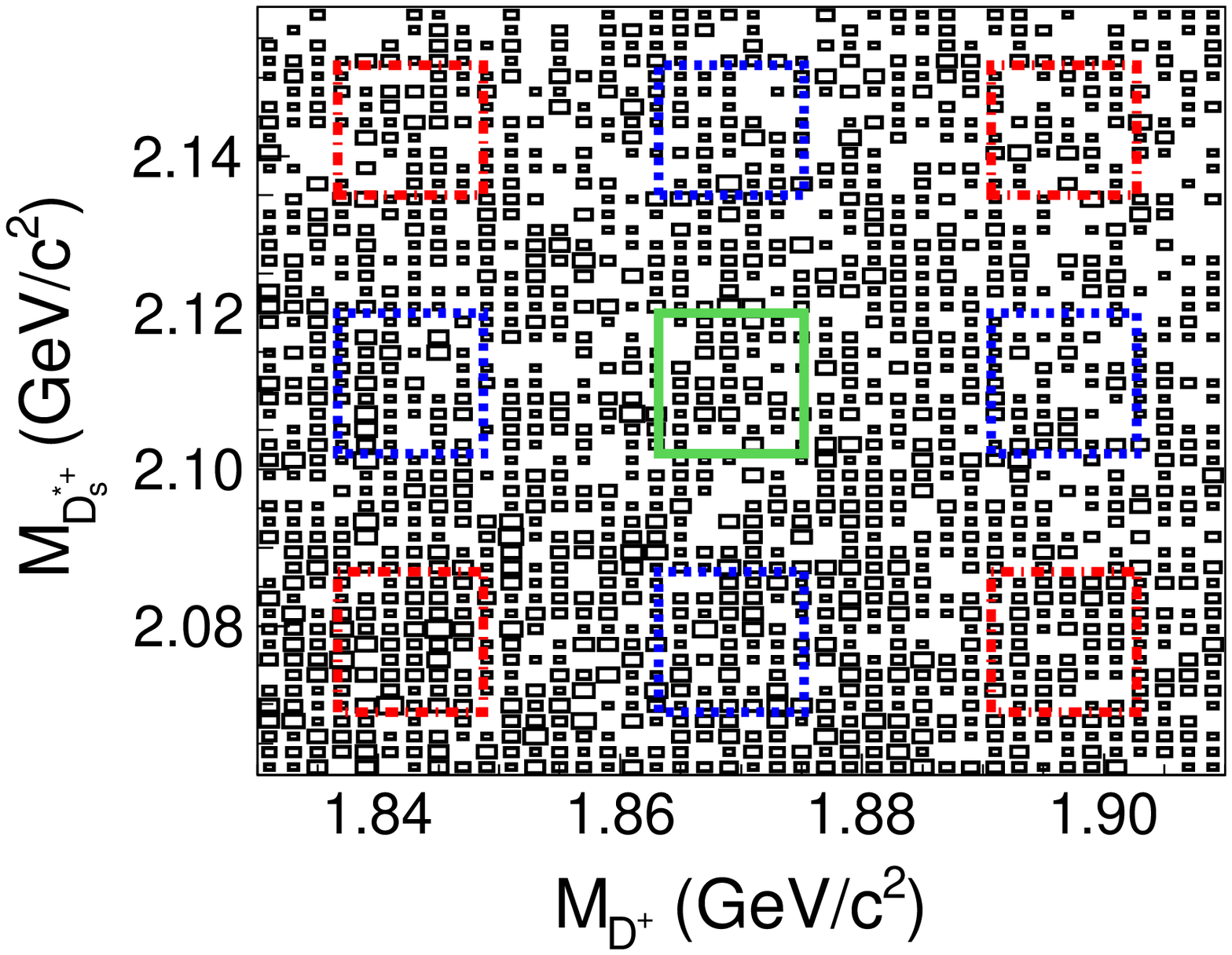}
		\includegraphics[height=6cm,width=8cm]{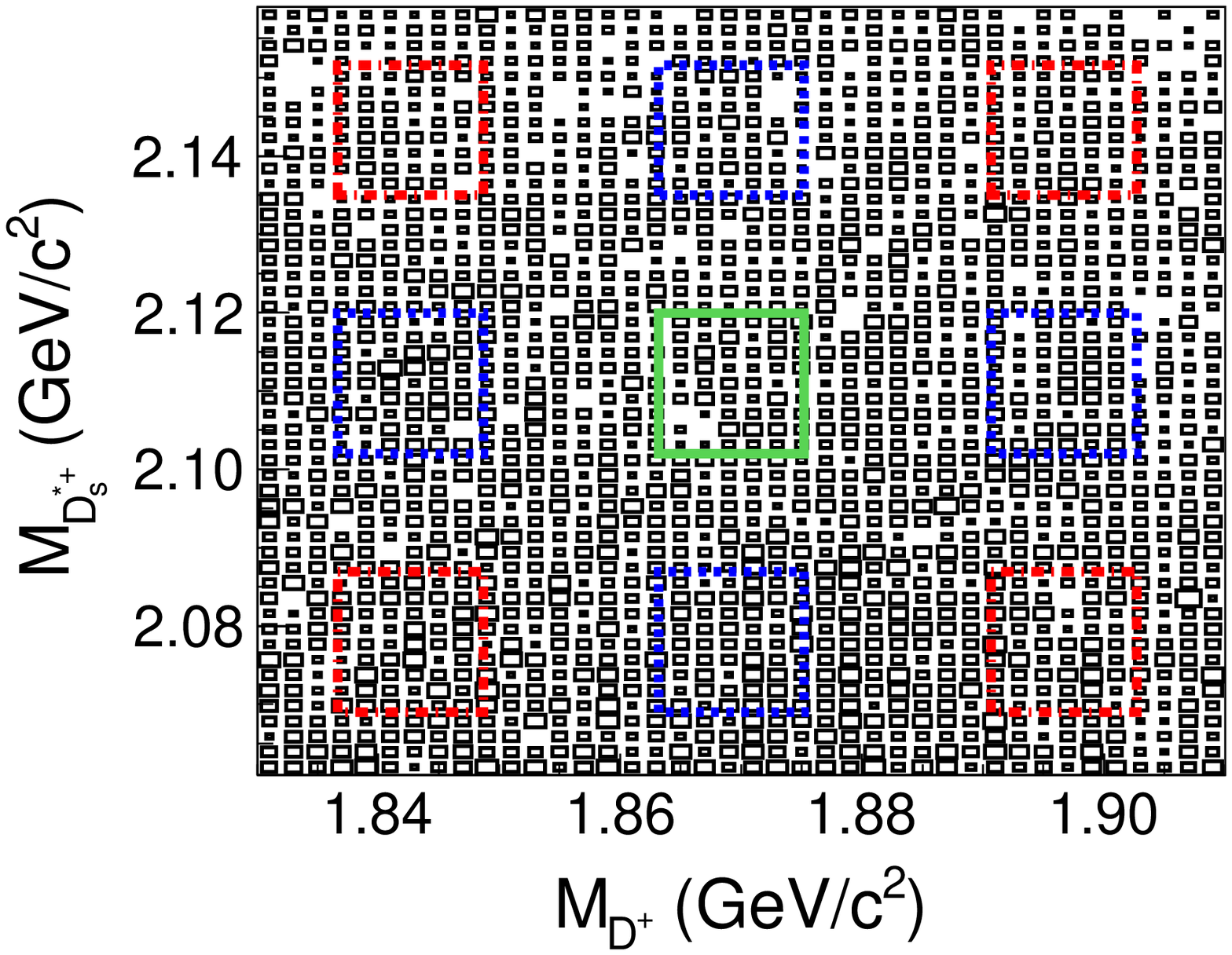}
        \put(-435,155){\bf (a)} \put(-205,155){\bf (b)}
		\caption{The scatter plots of $M_{D_{s}^{*+}}$ versus $M_{D^{+}}$ from (a) $\ones$ and (b) $\twos$ data samples.
		 The central solid boxes define the signal regions, and the red dash-dotted and blue dashed boxes
		 show the $M_{D^{+}}$ and $M_{D_{s}^{*+}}$ sideband regions described in the text.
        }\label{1s2s_2d}
	\end{center}
\end{figure*}

Figure~\ref{1s2s_2d} shows the scatter plots of $M_{D_{s}^{*+}}$ versus $M_{D^{+}}$ from $\ones$ and $\twos$ data samples, respectively. The central solid boxes show the signal regions of $D^{+}$ and $D_{s}^{*+}$. To check possible peaking backgrounds, the $M_{D^{+}}$ and $M_{D_{s}^{*+}}$ sidebands are selected, represented by the blue dashed (the total number of sideband events is denoted as $N_1$) and red dash-dotted boxes (the total number of sideband events is denoted as $N_2$) in Fig.~\ref{1s2s_2d}. The background contribution from the normalized $M_{D^{+}}$ and $M_{D_{s}^{*+}}$ sidebands is estimated to be $0.5 \times N_1 - 0.25 \times N_2$.

\begin{figure*}[htbp]
	\begin{center}
		\includegraphics[width=8cm]{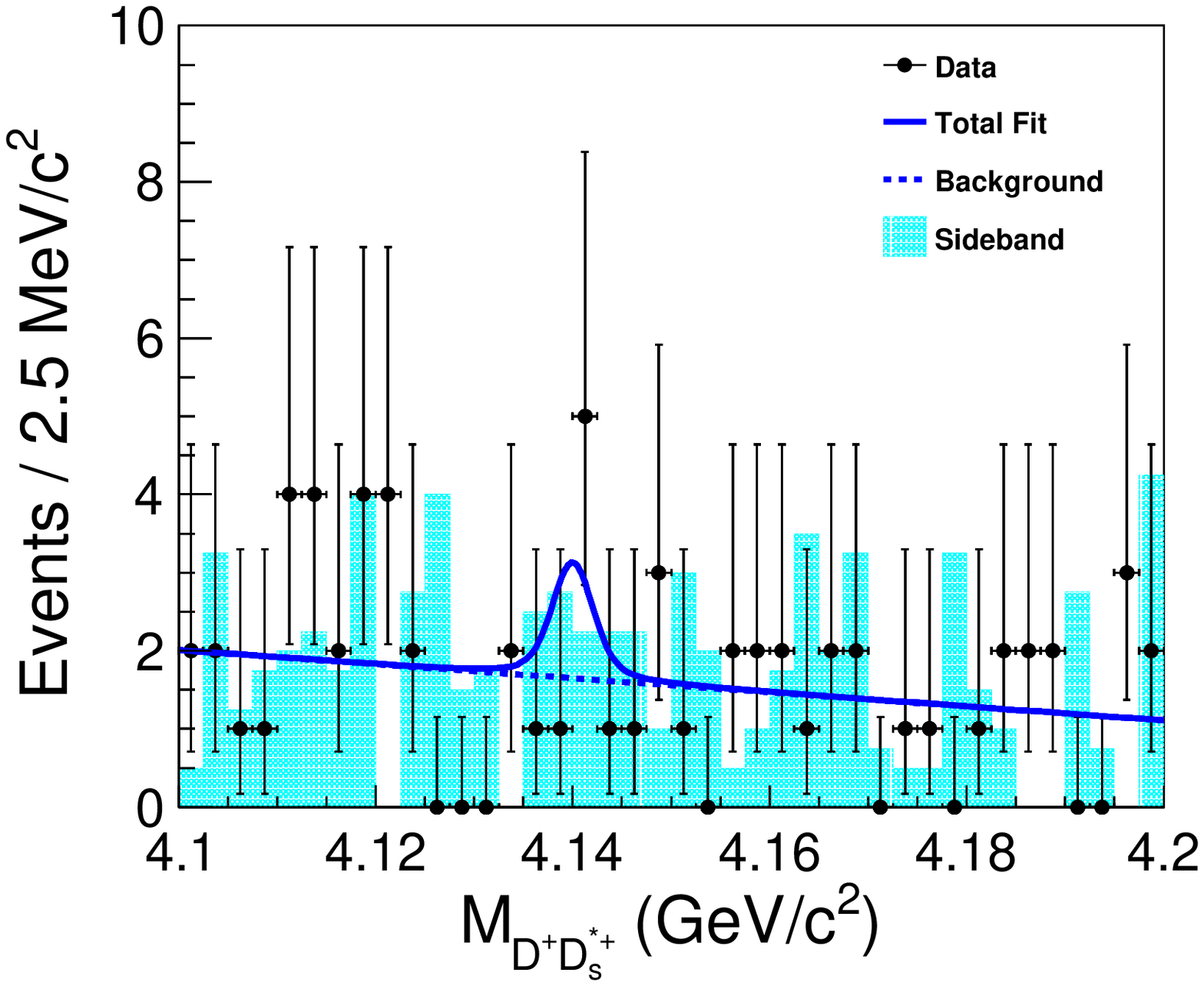}
		\includegraphics[width=8cm]{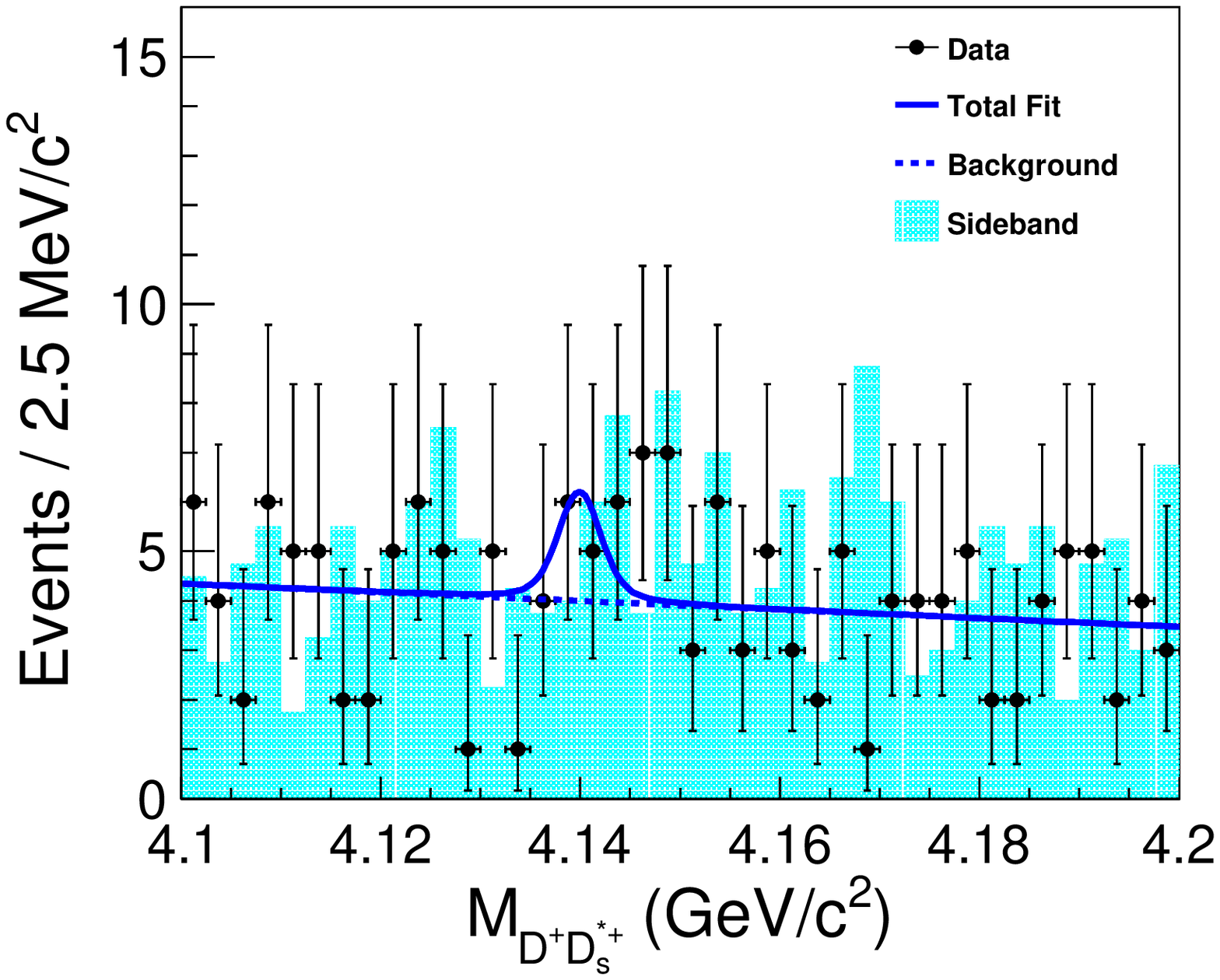}
        \put(-400,145){\bf (a)} \put(-170,145){\bf (b)}
		\caption{The invariant-mass spectra of $D^{+}D_{s}^{*+}$ in the (a) $\ones$ and (b) $\twos$ data samples. The cyan shaded histograms are from the normalized $M_{D^+}$ and $M_{D_{s}^{*+}}$ sideband events.
			The blue solid curves show the fitted results with the
			$\R$ mass fixed at 4.14~GeV/$c^2$ and width fixed at 2~MeV, and the blue dashed
			curves are the fitted backgrounds.
		}\label{DDs_mass_12s}
	\end{center}
\end{figure*}

Figure~\ref{DDs_mass_12s} shows the invariant-mass distributions of $D^{+}D_{s}^{*+}$
in the $\ones$ and $\twos$ data samples, together with the backgrounds from the normalized
$M_{D^{+}}$ and $M_{D_{s}^{*+}}$ sidebands. There are no evident signals for $\R$ states at the expected masses.
An unbinned extended maximum-likelihood fit repeated with $M_{\R}$ from 4.13 to 4.17~GeV/c$^2$ in steps
of 2.5~MeV/$c^2$, and $\Gamma_{R^{++}}$ from 0 to 5 MeV in steps of 1~MeV is performed to the
$M_{D^{+}D_{s}^{*+}}$ distribution. The signal shapes of $\R$ are described by a Gaussian function
($\Gamma_{R^{++}}$ = 0) or Breit-Wigner (BW) functions convolved with Gaussian functions
($\Gamma_{R^{++}}$ $\not=$ 0), where the parameters are fixed to those obtained from the fits to the corresponding MC simulated distributions.
The mass resolution of the $M_{D^{+}D_{s}^{*+}}$ is (1.7 $\pm$ 0.1)~MeV/$c^2$. There are no peaking backgrounds  found in the $M_{D^{+}}$ and $M_{D_{s}^{*+}}$ sidebands or in the $\onetwos$ inclusive MC samples~\cite{zhou}, so first-order polynomial functions with free parameters are taken as background shapes. The fitted results with the $\R$ mass fixed at 4.14~GeV/$c^2$ and width fixed at 2~MeV are shown in Fig.~\ref{DDs_mass_12s} as an example.
Assuming a Gaussian shape of the likelihoods, the local $\R$ significance is calculated using $\sqrt{-2\ln(\mathcal{L}_{0}/\mathcal{L}_{\rm max})}$, where $\mathcal{L}_{0}$ and $\mathcal{L}_{\rm max}$ are the likelihoods of the fits without and with a signal component, respectively. The fitted $\R$ signal yields at typically assumed mass points with $\Gamma_{R^{++}}$ fixed at values ranging from 0 to 5~MeV in steps of 1~MeV and the corresponding statistical significances are listed in Table~\ref{tab:1s2s}.

The branching fraction, $\BR(\onetwos \to R^{++} + anything) \times \BR(R^{++} \to D^{+} D_{s}^{*+})$, is calculated using
$$\frac{N^{\rm fit}}{N_{\onetwos} \times \sum_{i}\varepsilon_{i}\BR_{i}},$$
where $N^{\rm fit}$ is the fitted number of signal events,
$N_{\ones}$ = 1.02$\times10^{8}$ and $N_{\twos}$ = 1.58$\times10^{8}$ are the total numbers of $\ones$ and $\twos$ events,
the index $i$ runs for all final-state modes with $\varepsilon_i$ being the corresponding efficiency and ${\cal B}_i$
 the product of all secondary branching fractions of the mode $i$ [$\BR_{1}=\BR(D^+ \to K^{-} \pip \pip)\BR(D_{s}^{*+} \to D_{s}^{+}\gamma)
\BR(D_{s}^+ \to \phi(\to K^{+}K^{-})\pip)$, $\BR_{2}=\BR(D^+ \to K_{S}^{0} \pip)\BR(K_{S}^0 \to \pip \pim)
\BR(D_{s}^{*+} \to D_{s}^{+}\gamma)\BR(D_{s}^+ \to \phi(\to K^{+}K^{-})\pip)$, $\BR_{3}=\BR(D^+ \to K^{-}
\pip \pip)\BR(D_{s}^{*+} \to D_{s}^{+}\gamma)\BR(D_{s}^+ \to \bar{K}^{*}(892)^{0}(\to K^{-}\pip)K^{+})$,
$\BR_{4}=\BR(D^+ \to K_{S}^{0} \pip)\BR(K_{S}^0 \to \pip \pim) \BR(D_{s}^{*+} \to D_{s}^{+}\gamma)
\BR(D_{s}^+ \to \bar{K}^{*}(892)^{0}(\to K^{-}\pip)K^{+})$]. The calculated values
of $\BR(\onetwos \to R^{++} + anything) \times \BR(R^{++} \to D^{+} D_{s}^{*+})$
at typically assumed mass points are listed in Table~\ref{tab:1s2s}.

\begin{figure*}[htbp]
	\begin{center}
		\includegraphics[width=8cm]{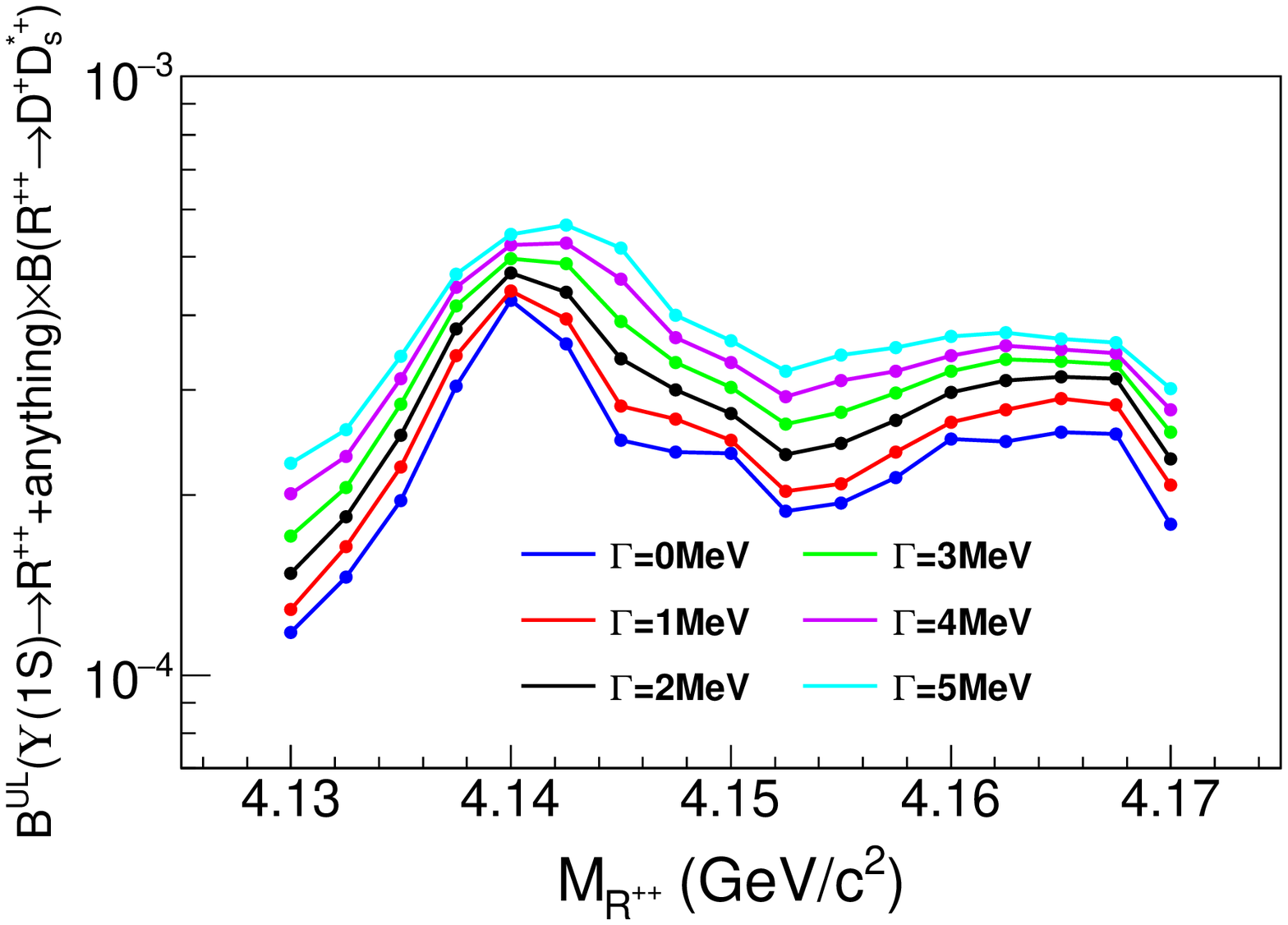}
		\includegraphics[width=8cm]{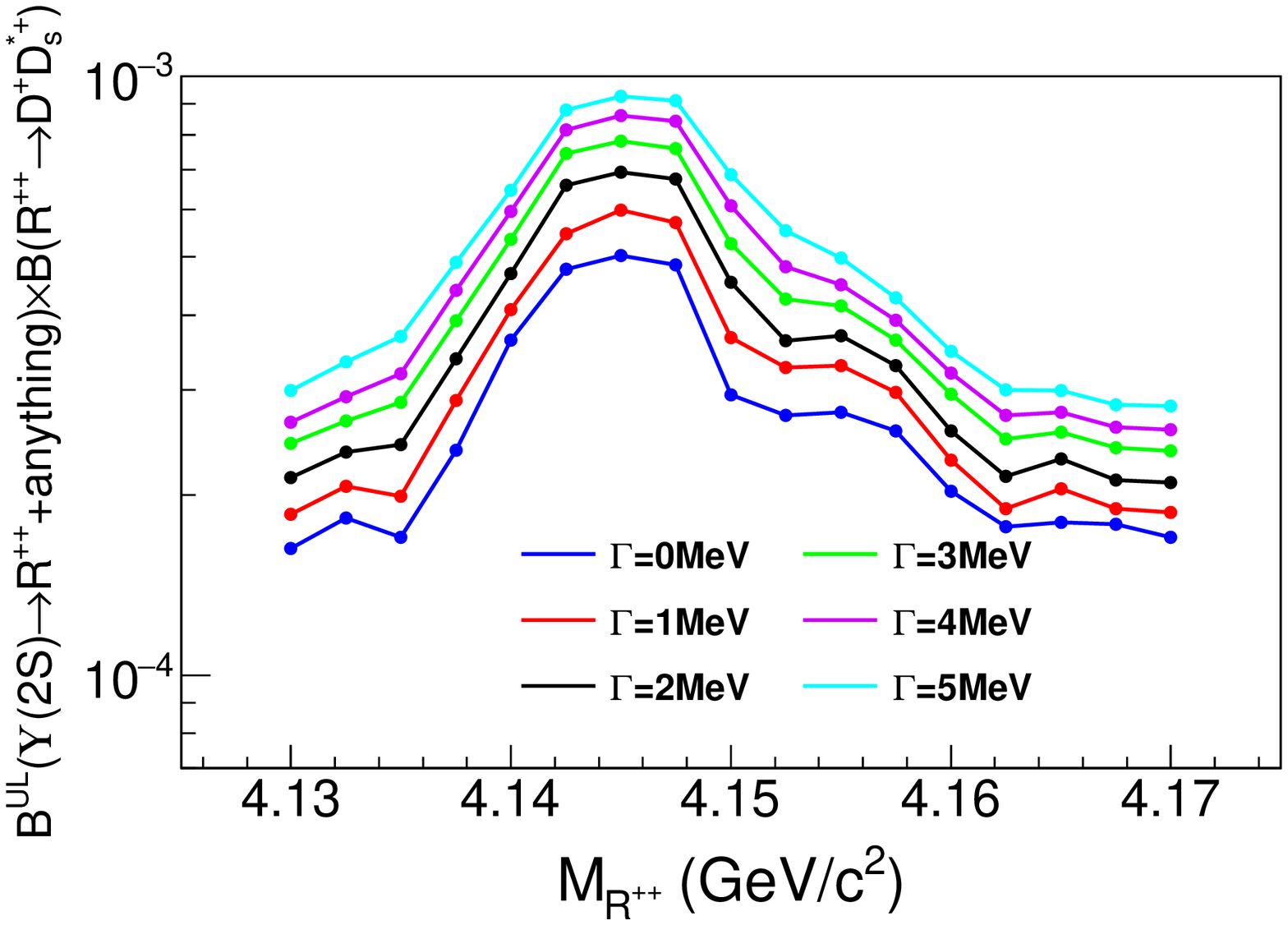}
        \put(-400,130){\bf (a)} \put(-170,130){\bf (b)}
		\caption{The 90\% C.L. upper limits on  (a) $\BR(\ones \to \R + anything) \times \BR(\R \to D^{+} D_{s}^{*+})$
			and (b) $\BR(\twos \to \R + anything) \times \BR(\R \to D^{+} D_{s}^{*+})$ as a function of the assumed $\R$ masses with widths
			varying from 0 to 5~MeV in steps of 1~MeV.}\label{1s2s_up}
	\end{center}
\end{figure*}

Since the statistical significance in each case is less than 3$\sigma$, Bayesian upper limits at the 90\% credibility level (C.L.)
on the numbers of signal events $(N^{\rm UL})$
assuming it follows a Poisson distribution with a  uniform prior probability density function are determined by solving the equation
$\int_0^{N^{\rm UL}} \mathcal{L} (x) dx /\int_0^{+\infty}\mathcal{L} (x) dx = 0.9$,
where $x$ is the number of fitted signal events and $\mathcal{L} (x)$ is the likelihood
function in the fit to data. Taking into account the systematic uncertainties
discussed below, the likelihood curve is convolved with a Gaussian function whose width equals the
corresponding total multiplicative systematic uncertainty. The calculated 90\% C.L. upper limits on the numbers of signal events
and the product branching fractions $(\BR^{\rm UL}(\onetwos \to R^{++} + anything) \times \BR(R^{++} \to D^{+} D_{s}^{*+}))$
in $\ones$ and $\twos$ inclusive decays at typically assumed mass points with
width fixed at values ranging from 0 to 5~MeV are listed in Table~\ref{tab:1s2s}.
The 90\% C.L. upper limits on the product branching fractions for all hypothetical $\R$ masses with widths varying
from 0 to 5~MeV are graphically shown in Fig.~\ref{1s2s_up}.

\begin{table*}[htbp]
	\caption{\label{tab:1s2s}
	Summary of the 90\% C.L. upper limits on the product branching fractions for $\onetwos \to \R + anything$
	with $\R \to D^{+}D_{s}^{*+}$ under typical assumptions of $\R$ mass ($M_{\R}$ in GeV/$c^2$) and width ($\Gamma_{R^{++}}$ in MeV) as examples,
    where $N^{\rm fit}$ is the number of fitted signal events, $N^{\rm UL}$ is the 90\% C.L. upper limit on the number of signal events
    taking into account systematic uncertainties, $\Sigma(\sigma)$ is the local $\R$ significance, $\Sigma_{i}(\epsilon_i\BR_i)$ is the sum of
	product of the detection efficiency and the product of all secondary branching fractions for each reconstruction mode,
	$\sigma_{\rm multi}$ is the total multiplicative systematic uncertainty, $\sigma_{\rm add}$ is the additive systematic uncertainty, $\BR$ ($\BR(\onetwos \to R^{++} + anything) \times \BR(R^{++} \to D^{+} D_{s}^{*+})$) is the product branching fraction for $\onetwos \to \R + anything$ with $\R \to D^{+}D_{s}^{*+}$, and $\BR^{\rm UL}$ ($\BR^{\rm UL}(\onetwos \to R^{++} + anything) \times \BR(R^{++} \to D^{+} D_{s}^{*+})$) is the 90\% C.L. upper limit on the product branching fraction with systematic uncertainties included.}
	\scriptsize
	\vspace{0.2cm}
	\begin{tabular}{ccc@{$/$}cc@{$/$}cc@{$/$}cccc@{$/$}cc@{$/$}cc}
	\hline\hline
	\multicolumn{15}{c}{$\ones$/$\twos$ $\to \R + anything$, $\R \to D^{+}D_{s}^{*+}$}\\\hline
	 $M_{\R}$ & $\Gamma_{R^{++}}$ & \multicolumn{2}{c}{$N^{\rm fit}$} & \multicolumn{2}{c}{$N^{\rm UL}$} & \multicolumn{2}{c}{ $\Sigma(\sigma)$} & $\Sigma_{i}(\epsilon_i\BR_i)$  ($\times10^{-5}$) & $\sigma_{\rm multi}(\%)$ & \multicolumn{2}{c}{$\sigma_{\rm add}(\%)$} & \multicolumn{2}{c}{ $\BR(\times 10^{-5})$} & $\BR^{\rm UL}(\times 10^{-5})$  \\
	\hline
4.13 & 0 & $-4.2$$\pm$3.7 & $-2.6$$\pm$2.7 &  2.7&5.4  &   -&-   & 22.4/21.0 & 8.0/8.1 &  3.5&5.9 & $-18.4\pm$16.2& $-7.8\pm$8.1    & 11.8/16.3  \\
4.13 & 1 & $-4.0$$\pm$3.9 & $-3.3$$\pm$3.3 &  2.9&6.1  &   -&-   & 22.1/20.8 & 8.0/8.1 &  3.8&6.2 & $-17.7\pm$17.3&$-10.0\pm$10.0   & 12.9/18.6  \\
4.13 & 2 & $-4.1$$\pm$4.3 & $-3.9$$\pm$3.8 &  3.3&6.9  &   -&-   & 21.9/20.4 & 8.0/8.1 &  6.5&7.8 & $-18.4\pm$19.2&$-12.1\pm$11.8   & 14.8/21.4  \\
4.13 & 3 & $-4.5$$\pm$4.8 & $-4.5$$\pm$4.3 &  3.8&7.7  &   -&-   & 21.8/20.0 & 8.0/8.1 & 11.8&8.9 & $-20.2\pm$21.6&$-14.2\pm$13.6   & 17.1/24.4  \\
4.13 & 4 & $-4.8$$\pm$5.2 & $-5.1$$\pm$4.9 &  4.4&8.5  &   -&-   & 21.5/20.3 & 8.0/8.1 & 12.8&9.0 & $-21.9\pm$23.7&$-15.9\pm$15.3   & 20.1/26.5  \\
4.13 & 5 & $-5.2$$\pm$5.8 & $-5.8$$\pm$5.6 &  5.0&9.5  &   -&-   & 21.7/20.1 & 8.0/8.1 & 15.9&9.2 & $-23.5\pm$26.2&$-18.3\pm$17.6   & 22.6/29.9  \\
4.14 & 0 &    3.7$\pm$2.9 &    4.3$\pm$4.0 &  9.7&12.0 & 1.6&1.2 & 22.5/20.9 & 8.0/8.1 &  7.6&8.6 &  16.1$\pm$12.6& 13.0$\pm$12.1   & 42.3/36.3  \\
4.14 & 1 &    3.7$\pm$3.0 &    4.9$\pm$4.5 &  9.9&13.4 & 1.5&1.2 & 22.1/20.8 & 8.0/8.1 &  7.9&9.7 &  16.4$\pm$13.3& 14.9$\pm$13.7   & 43.9/40.8  \\
4.14 & 2 &    3.7$\pm$3.2 &    5.6$\pm$5.1 & 10.5&15.2 & 1.3&1.2 & 21.9/20.5 & 8.0/8.1 &  9.8&12.2&  16.6$\pm$14.3& 17.3$\pm$15.7   & 47.0/46.9  \\
4.14 & 3 &    3.6$\pm$3.5 &    6.4$\pm$5.6 & 11.0&17.0 & 1.2&1.3 & 21.7/20.1 & 8.0/8.1 & 12.0&13.5&  16.3$\pm$15.8& 20.2$\pm$17.6   & 49.7/53.5  \\
4.14 & 4 &    3.5$\pm$3.7 &    7.2$\pm$6.3 & 11.5&19.0 & 1.0&1.3 & 21.5/20.2 & 8.0/8.1 & 14.7&14.7&  16.0$\pm$16.9& 22.6$\pm$19.7   & 52.4/59.5  \\
4.14 & 5 &    3.1$\pm$4.0 &    7.8$\pm$6.7 & 12.0&20.5 & 0.8&1.3 & 21.6/20.1 & 8.0/8.1 & 15.8&15.8&  14.1$\pm$18.2& 24.6$\pm$21.1   & 54.5/64.6  \\
4.15 & 0 &    0.0$\pm$2.1 &    2.2$\pm$3.6 &  5.4&9.7  &   -&0.6 & 22.5/20.9 & 8.0/8.1 &  3.7&13.2&   0.0$\pm$9.2 &  6.7$\pm$10.9   & 23.5/29.4  \\
4.15 & 1 & $-0.2$$\pm$2.3 &    3.2$\pm$4.5 &  5.6&12.0 &   -&0.8 & 22.2/20.7 & 8.0/8.1 &  3.8&14.8&  $-0.9\pm$10.2&  9.8$\pm$13.8   & 24.7/36.7  \\
4.15 & 2 & $-0.3$$\pm$2.6 &    4.7$\pm$5.2 &  6.1&14.7 &   -&1.0 & 21.8/20.5 & 8.0/8.1 &  5.2&13.3&  $-1.3\pm$11.7& 14.5$\pm$16.1   & 27.4/45.4  \\
4.15 & 3 & $-0.5$$\pm$2.8 &    5.9$\pm$5.9 &  6.7&16.8 &   -&1.1 & 21.7/20.2 & 8.0/8.1 &  6.8&11.3&  $-2.3\pm$12.7& 18.5$\pm$18.5   & 30.3/52.6  \\
4.15 & 4 & $-0.7$$\pm$3.1 &    7.5$\pm$6.4 &  7.3&19.4 &   -&1.3 & 21.5/20.2 & 8.0/8.1 &  9.0&9.3 &  $-3.2\pm$14.1& 23.5$\pm$20.1   & 33.3/60.8  \\
4.15 & 5 & $-1.0$$\pm$3.5 &    8.8$\pm$7.0 &  7.9&21.8 &   -&1.4 & 21.4/20.1 & 8.0/8.1 &  9.7&9.9 &  $-4.6\pm$16.0& 27.7$\pm$22.0   & 36.2/68.6  \\
4.16 & 0 &    1.0$\pm$2.1 & $-1.9$$\pm$3.4 &  5.7&6.7  & 0.5&-   & 22.5/20.9 & 8.0/8.1 &  3.9&6.3 &   4.4$\pm$9.2 & $-5.8\pm$10.3   & 24.8/20.3  \\
4.16 & 1 &    0.9$\pm$2.3 & $-1.6$$\pm$3.8 &  6.0&7.5  & 0.4&-   & 22.2/20.7 & 8.0/8.1 &  5.2&5.6 &   4.0$\pm$10.2& $-4.9\pm$11.6   & 26.5/22.9  \\
4.16 & 2 &    0.7$\pm$2.7 & $-1.6$$\pm$4.3 &  6.6&8.3  & 0.3&-   & 21.8/20.5 & 8.0/8.1 &  5.3&5.1 &   3.1$\pm$12.1& $-4.9\pm$13.3   & 29.7/25.6  \\
4.16 & 3 &    0.6$\pm$3.0 & $-1.5$$\pm$4.8 &  7.1&9.4  & 0.2&-   & 21.6/20.2 & 8.0/8.1 &  6.2&5.6 &   2.7$\pm$13.6& $-4.7\pm$15.0   & 32.2/29.5  \\
4.16 & 4 &    0.6$\pm$3.2 & $-1.7$$\pm$5.3 &  7.5&10.2 & 0.2&-   & 21.5/20.2 & 8.0/8.1 &  2.3&6.2 &   2.7$\pm$14.6& $-5.3\pm$16.6   & 34.2/32.0  \\
4.16 & 5 &    0.5$\pm$3.5 & $-1.7$$\pm$5.7 &  8.0&11.0 & 0.1&-   & 21.3/20.0 & 8.0/8.1 &  3.1&5.8 &   2.3$\pm$16.1& $-5.4\pm$18.0   & 36.8/34.8  \\
4.17 & 0 & $-2.9$$\pm$2.0 & $-2.1$$\pm$2.8 &  4.1&5.6  &   -&-   & 22.5/20.8 & 8.0/8.1 &  5.8&7.3 & $-12.6\pm$8.7 & $-6.4\pm$8.5    & 17.9/17.0  \\
4.17 & 1 & $-2.4$$\pm$2.2 & $-2.6$$\pm$3.2 &  4.7&6.1  &   -&-   & 22.2/20.6 & 8.0/8.1 &  6.2&7.8 & $-10.6\pm$9.7 & $-8.0\pm$9.8    & 20.8/18.7  \\
4.17 & 2 & $-2.5$$\pm$2.5 & $-3.2$$\pm$3.7 &  5.1&6.8  &   -&-   & 21.7/20.5 & 8.0/8.1 &  6.4&9.1 & $-11.3\pm$11.3& $-9.9\pm$11.4   & 23.0/21.0  \\
4.17 & 3 & $-2.3$$\pm$2.8 & $-3.9$$\pm$4.4 &  5.6&7.6  &   -&-   & 21.5/20.3 & 8.0/8.1 &  6.7&12.4& $-10.5\pm$12.8&$-12.2\pm$13.7   & 25.5/23.7  \\
4.17 & 4 & $-2.4$$\pm$3.1 & $-4.4$$\pm$4.8 &  6.1&8.2  &   -&-   & 21.5/20.2 & 8.0/8.1 &  7.0&13.3& $-10.9\pm$14.1&$-13.8\pm$15.0   & 27.8/25.7  \\
4.17 & 5 & $-2.6$$\pm$3.3 & $-5.0$$\pm$5.4 &  6.5&8.9  &   -&-   & 21.2/20.0 & 8.0/8.1 &  6.8&14.5& $-12.0\pm$15.3&$-15.8\pm$17.1   & 30.1/28.2  \\	
	\hline\hline
	\end{tabular}
\end{table*}

\section{\boldmath $\EE \to R^{++}$ + anything at $\sqrt{s}$ = 10.520, 10.580, and 10.867~$\textrm{GeV}$}
\begin{figure*}[htbp]
	\begin{center}
		\includegraphics[width=5.9cm]{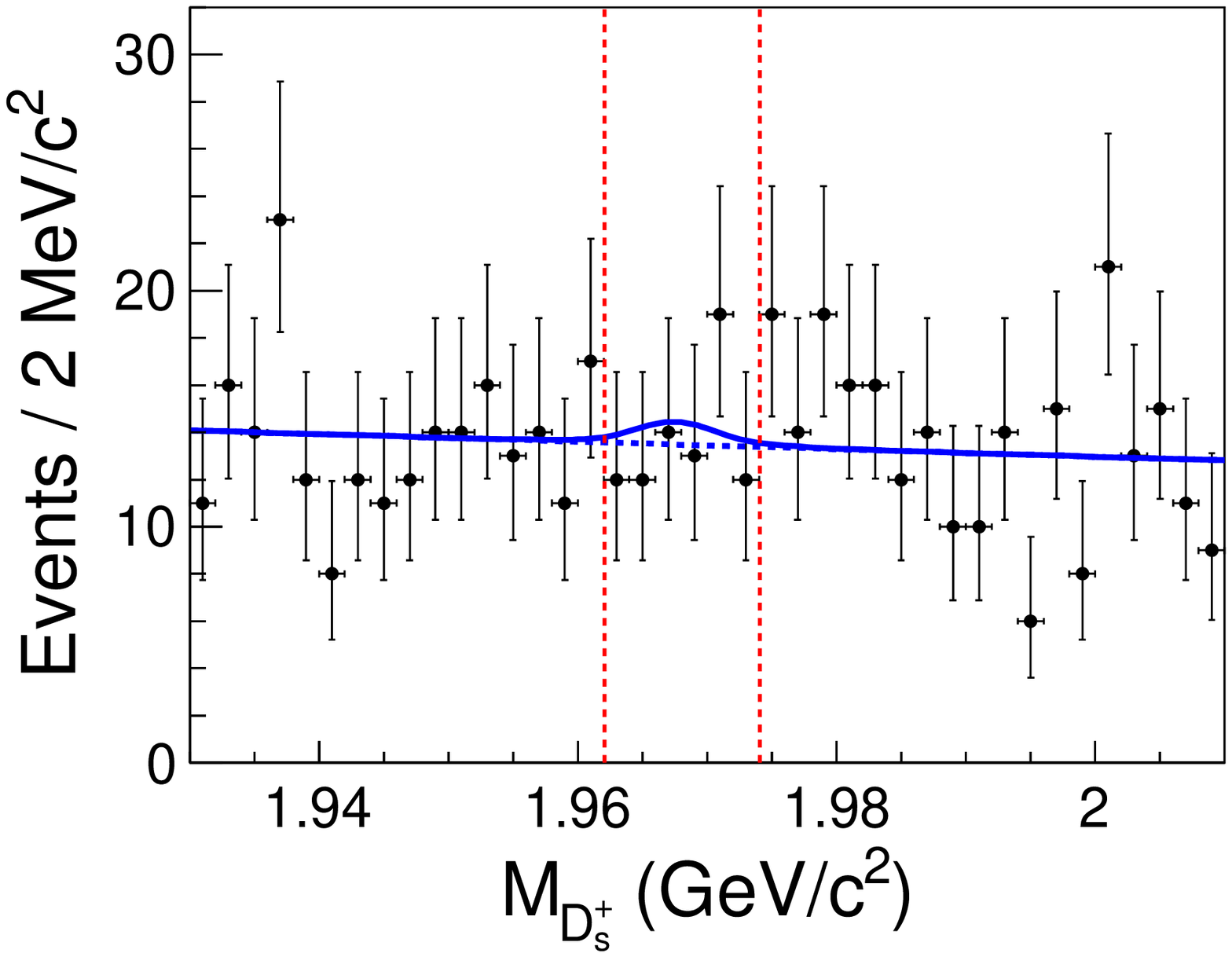}
		\includegraphics[width=5.9cm]{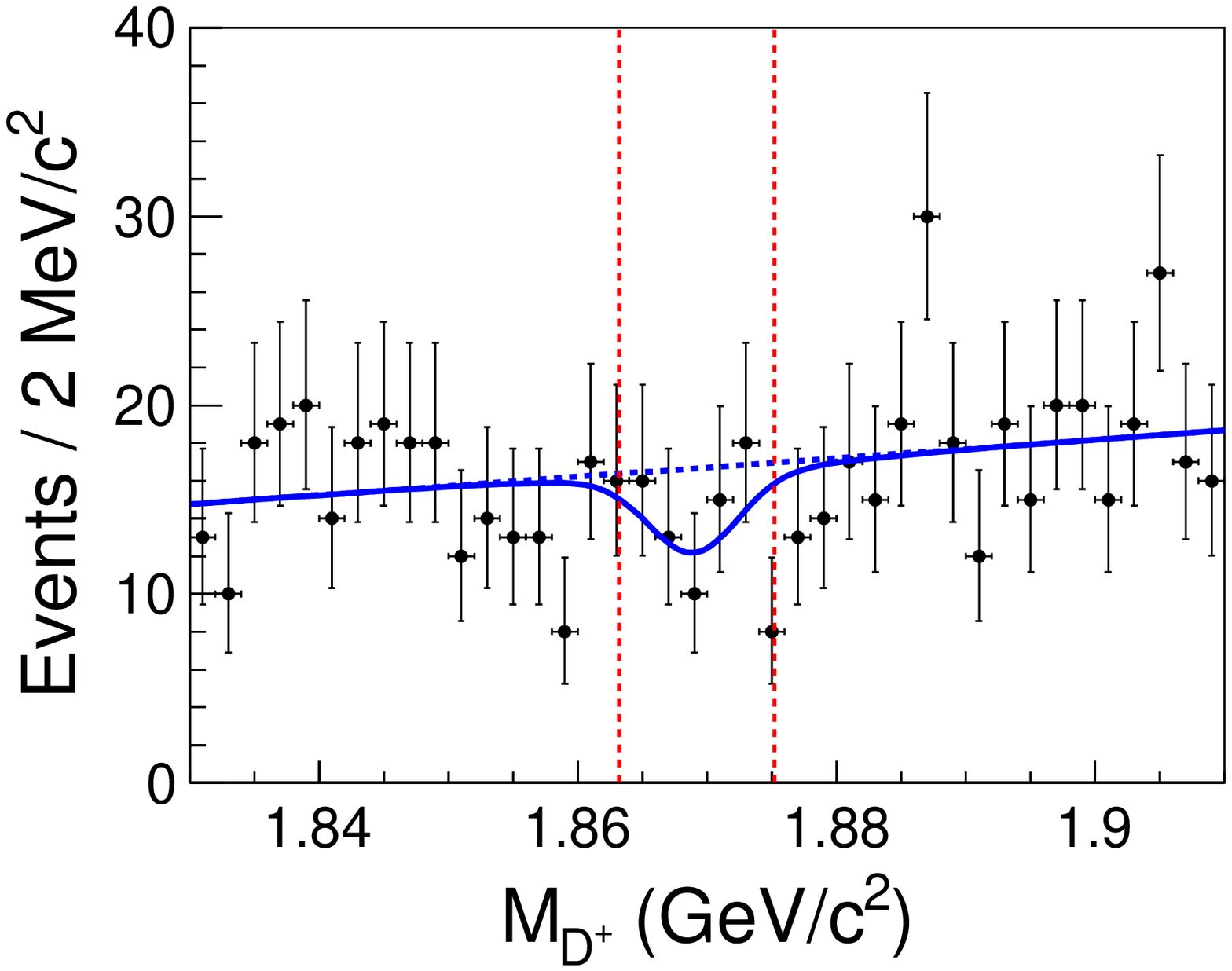}
		\includegraphics[width=5.9cm]{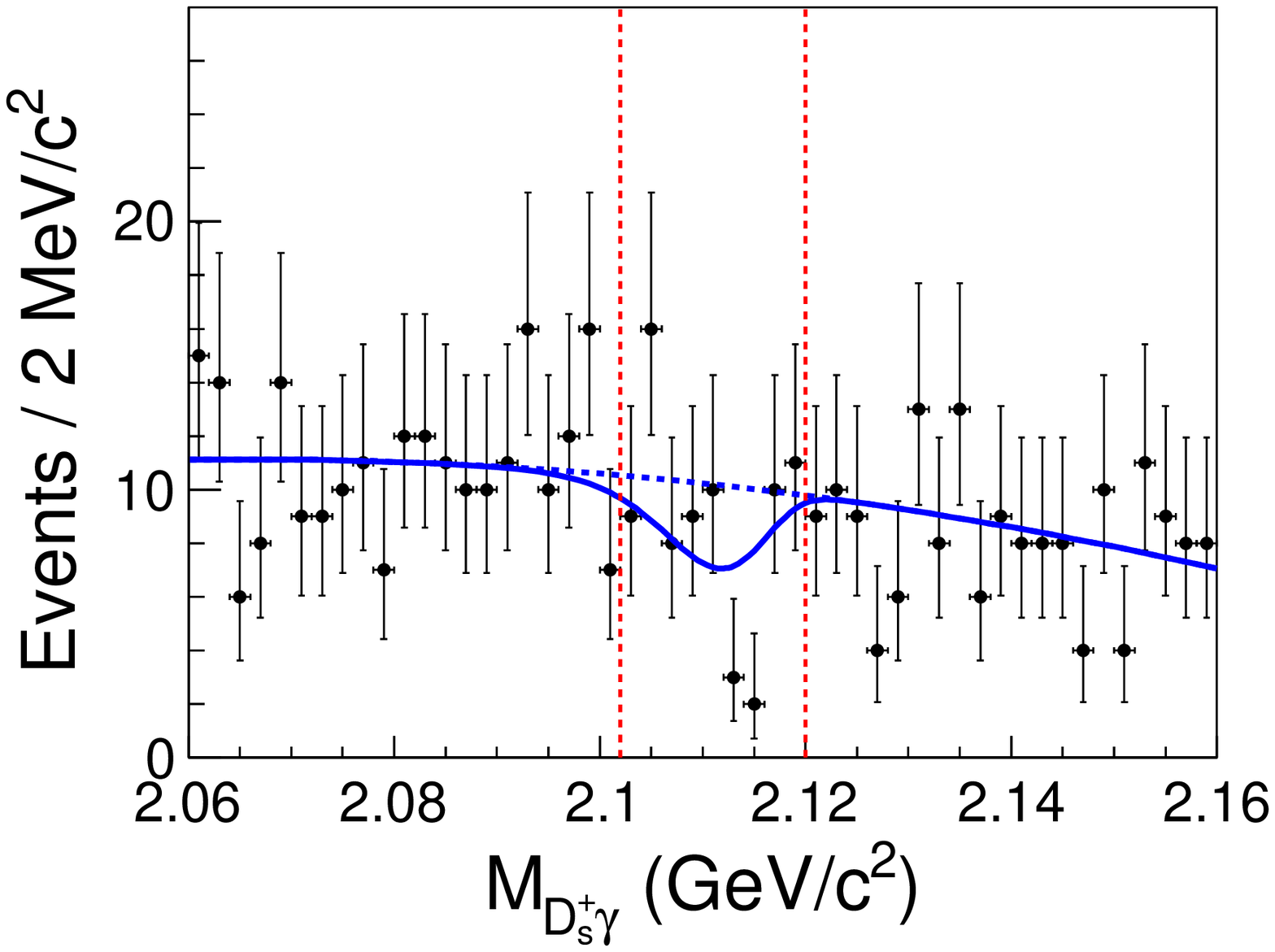}
		\put(-380,100){\bf (a)} \put(-210,100){\bf (b)}  \put(-38,100){\bf (c)}
		\caption{The invariant-mass spectra of the (a) $D_{s}^{+}$, (b) $D^{+}$, and (c) $D_{s}^{*+}$ candidates summed over four
			reconstructed modes from $\sqrt{s}$ = 10.520~GeV data. The points with error bars represent the data,
			the solid curves show the results of the best fits to the data, and the blue dashed curves are the fitted backgrounds.
			The red dashed lines show the required signal regions.
		}\label{int_mass_off}
	\end{center}
\end{figure*}

\begin{figure*}[htbp]
	\begin{center}
		\includegraphics[width=5.9cm]{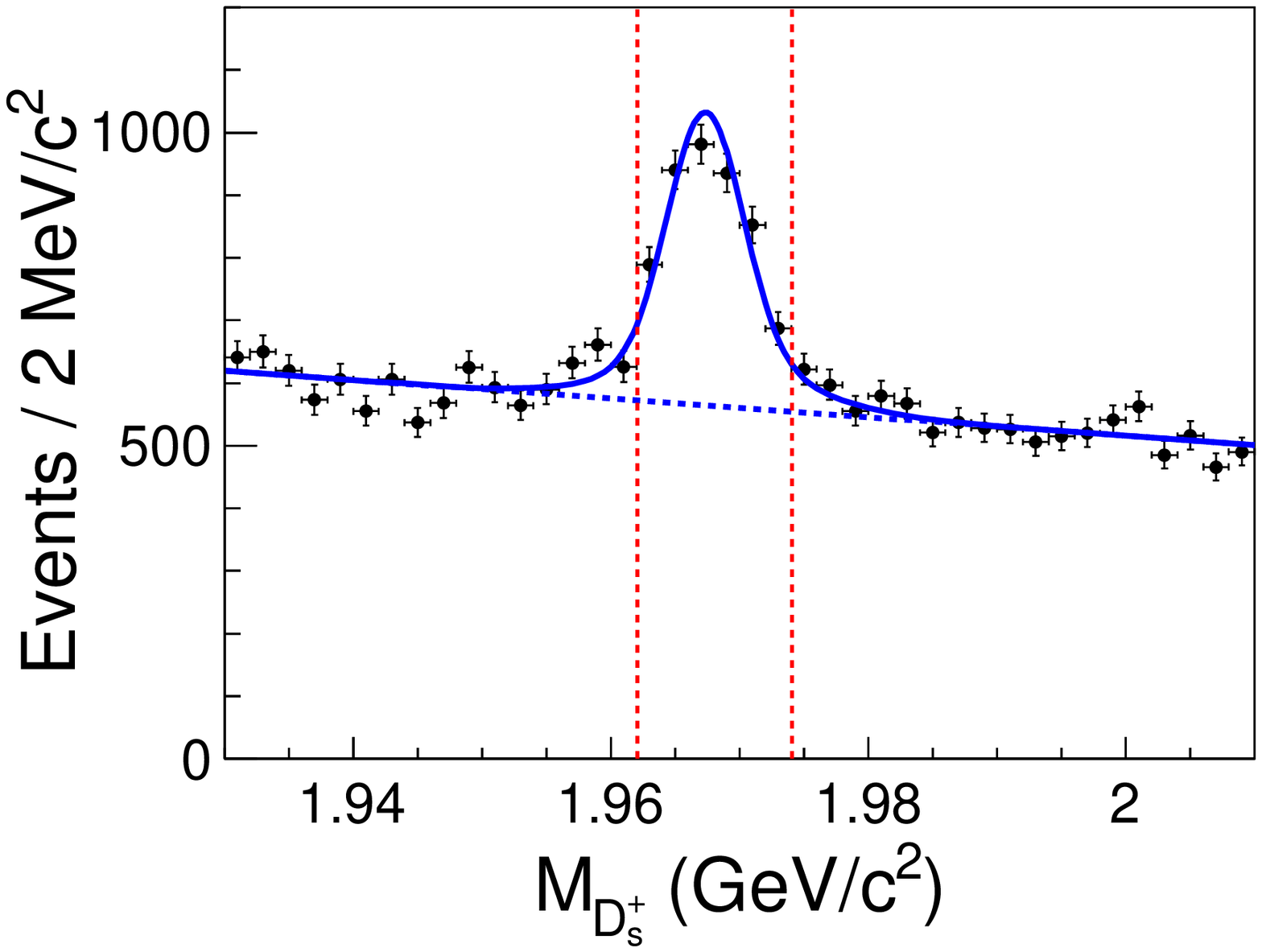}
		\includegraphics[width=5.9cm]{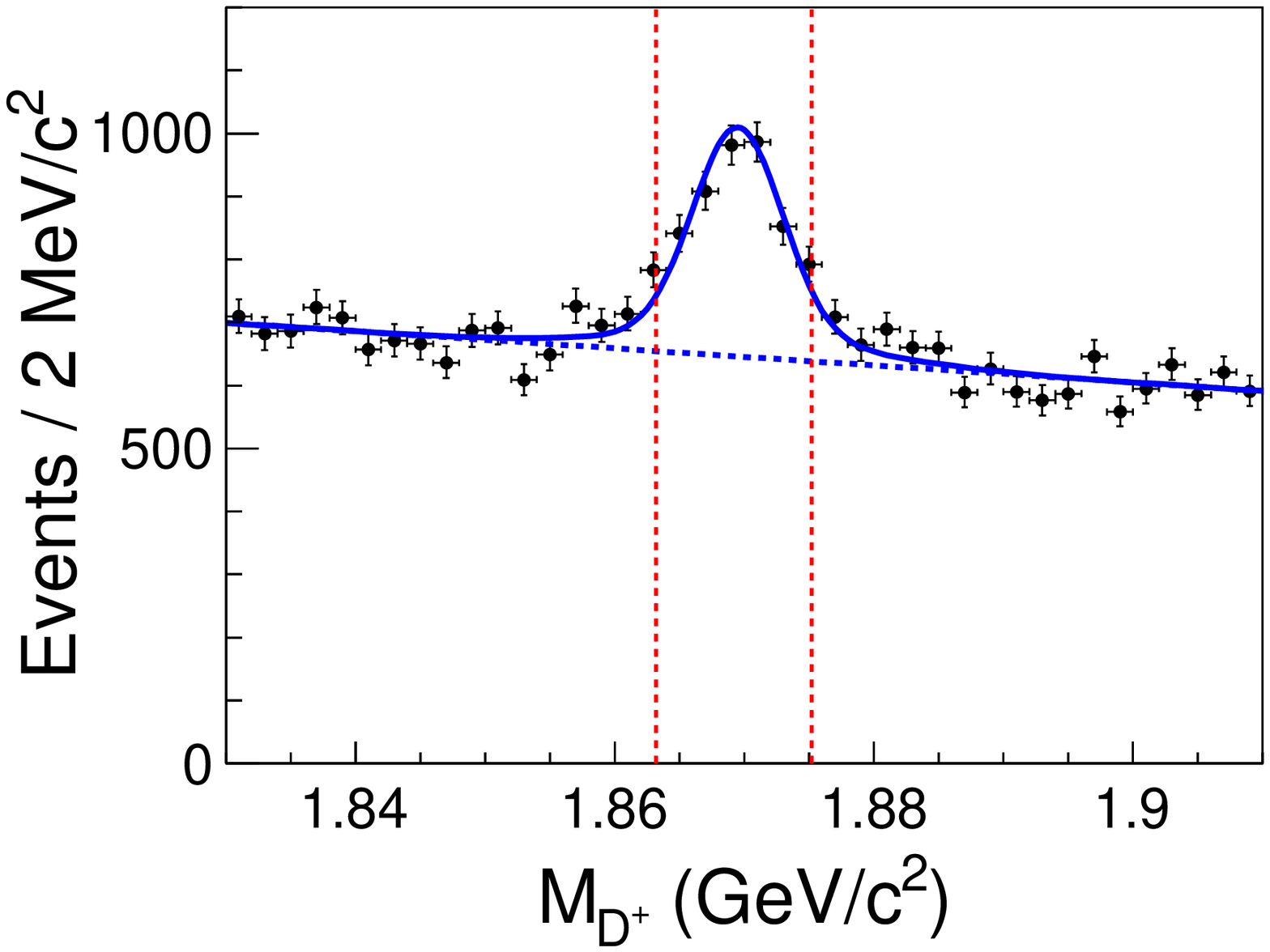}
		\includegraphics[width=5.9cm]{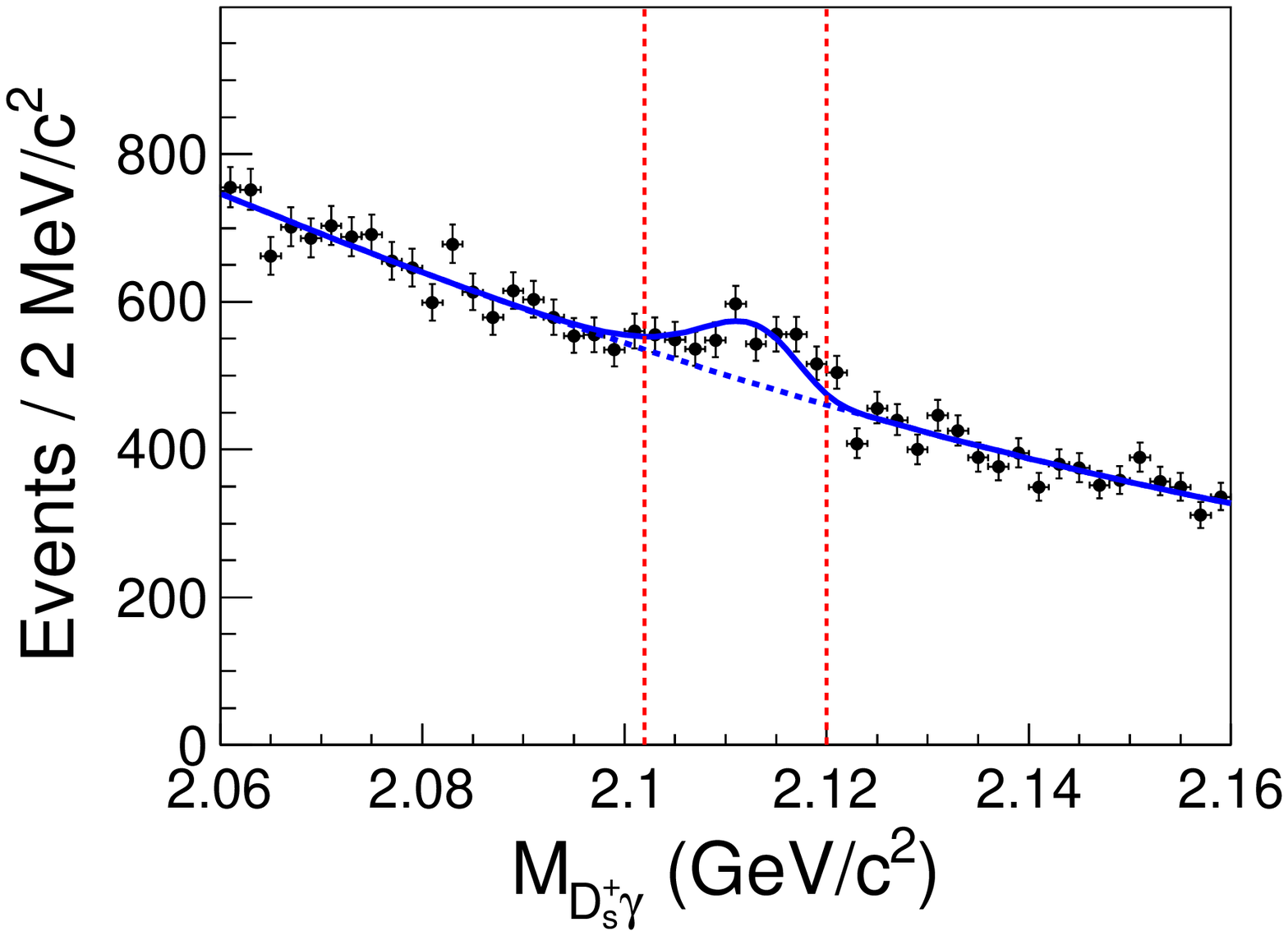}
			\put(-380,100){\bf (a)} \put(-210,100){\bf (b)}  \put(-38,100){\bf (c)}
		\caption{The invariant-mass spectra of the (a) $D_{s}^{+}$, (b) $D^{+}$, and (c) $D_{s}^{*+}$ candidates summed over four
			reconstructed modes from $\sqrt{s}$ = 10.580~GeV data. The points with error bars represent the data,
			the solid curves show the results of the best fits to the data, and the blue dashed curves are the fitted backgrounds.
			The red dashed lines show the required signal regions.
		}\label{int_mass_4s}
	\end{center}
\end{figure*}

\begin{figure*}[htbp]
	\begin{center}
		\includegraphics[width=5.9cm]{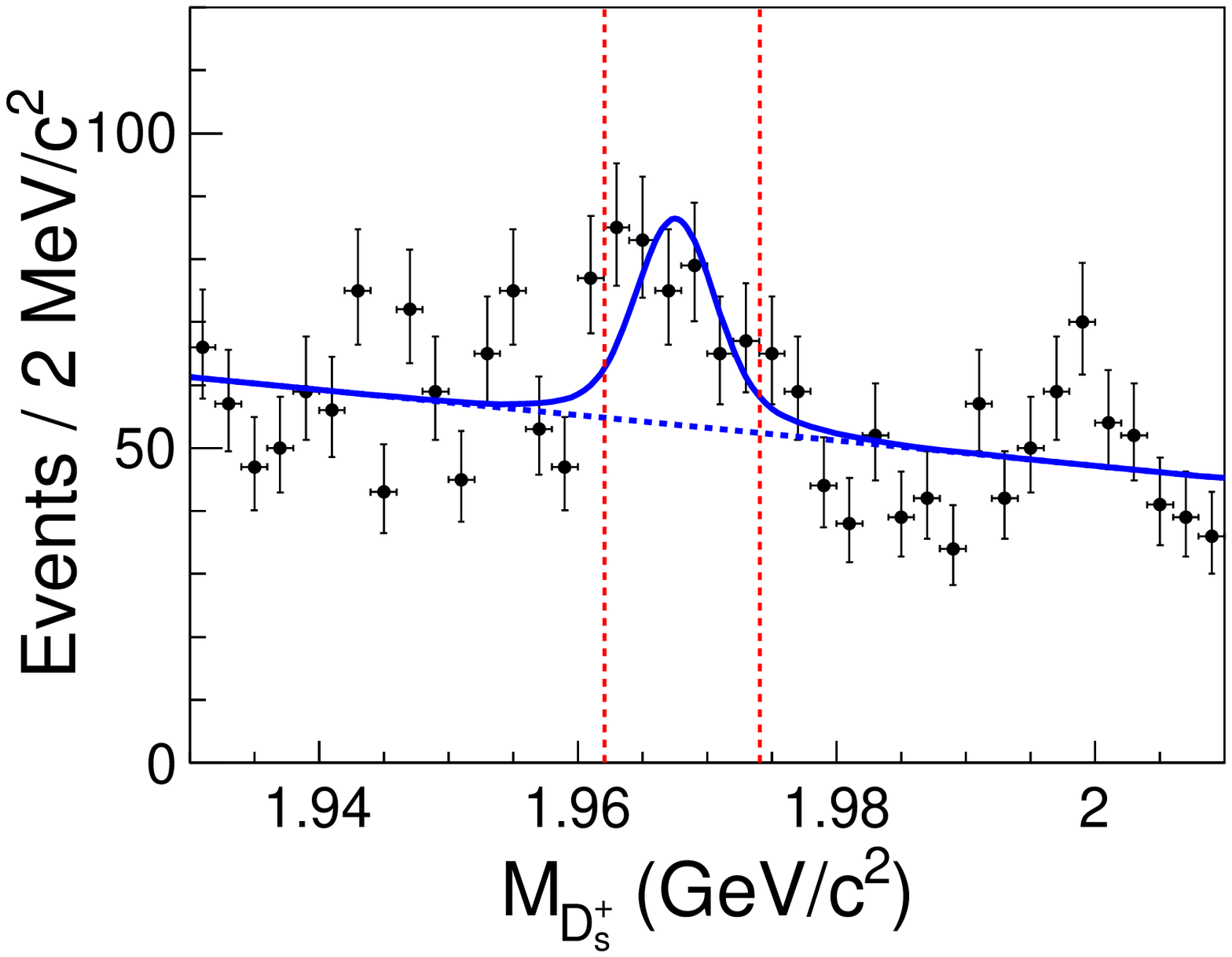}
		\includegraphics[width=5.9cm]{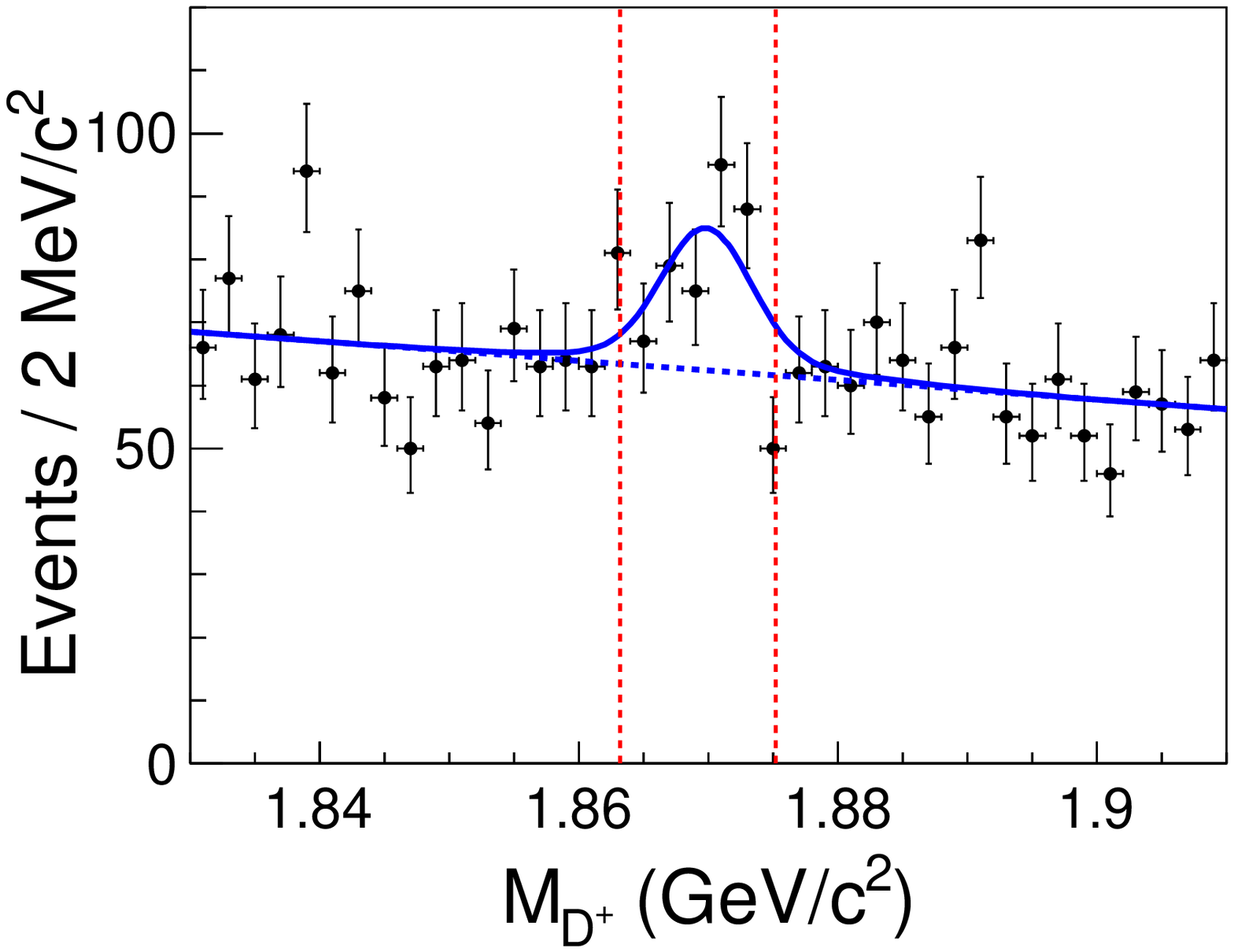}
		\includegraphics[width=5.9cm]{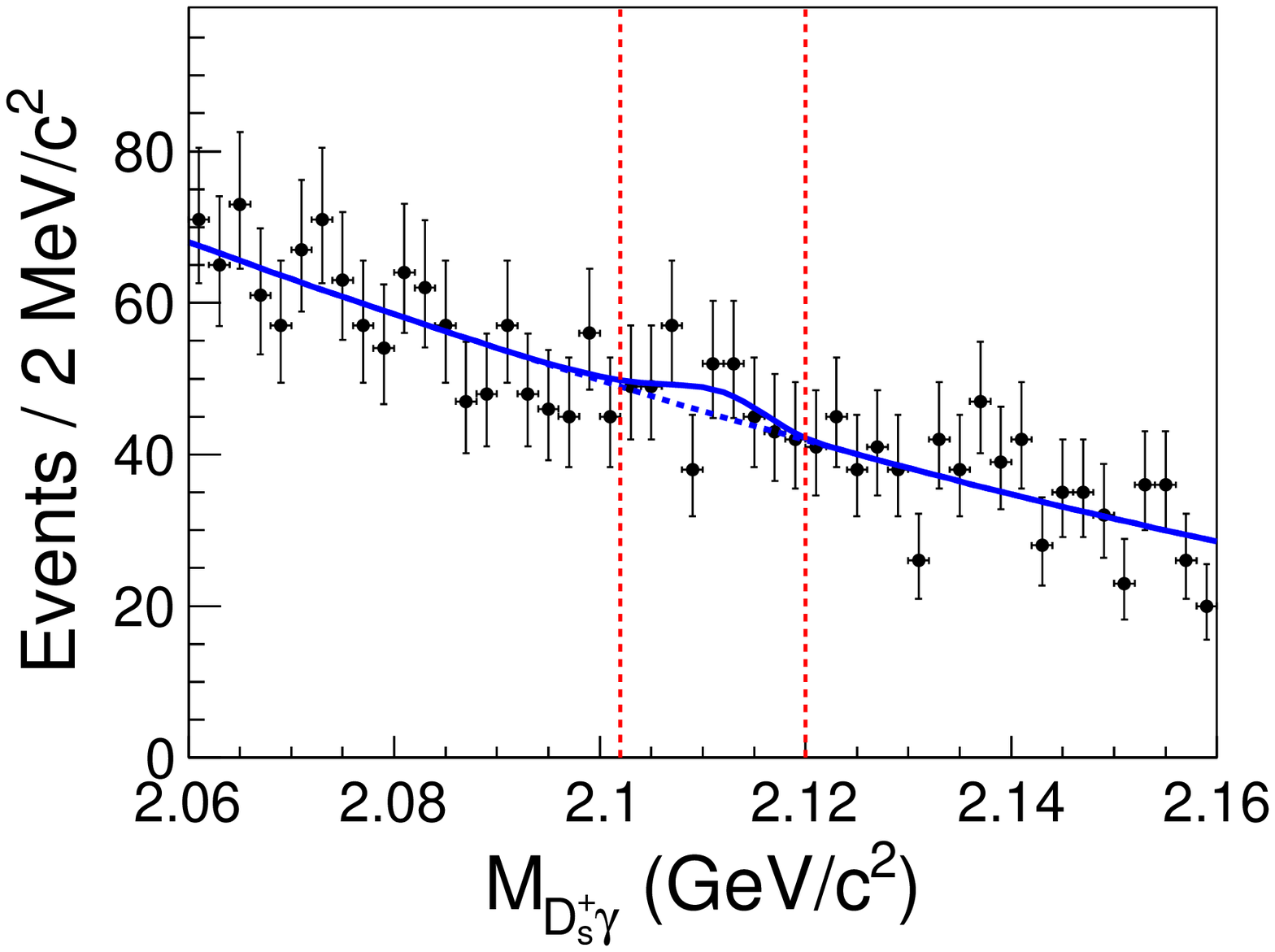}
		\put(-380,100){\bf (a)} \put(-210,100){\bf (b)}  \put(-38,100){\bf (c)}
		\caption{The invariant-mass spectra of the (a) $D_{s}^{+}$, (b) $D^{+}$, and (c) $D_{s}^{*+}$ candidates summed over four
			reconstructed modes from $\sqrt{s}$ = 10.867~GeV data. The points with error bars represent the data,
			the solid curves show the results of the best fits to the data, and the blue dashed curves are the fitted backgrounds.
			The red dashed lines show the required signal regions.
		}\label{int_mass_5s}
	\end{center}
\end{figure*}

In this section, we search for the doubly-charged $DDK$ bound state via direct production in $\EE$ collisions at $\sqrt{s}$ = 10.520, 10.580, and 10.867~GeV. After the application of the selection criteria, the invariant-mass distributions of $D_{s}^{+}$, $D^{+}$,
and $D_{s}^{*+}$ candidates from $\sqrt{s}$ = 10.520, 10.580, and 10.867~GeV data samples are shown in Figs.~\ref{int_mass_off},~\ref{int_mass_4s}, and~\ref{int_mass_5s}, respectively, together with results of the fits. When drawing each distribution, the signal mass windows of other intermediate states are required. Since the $\sqrt{s}$=10.520~GeV data sample is below the $B_{(s)}\bar{B}_{(s)}$ threshold, there are no $D_{s}^{+}$, $D^{+}$, or $D_{s}^{*+}$ candidates from the $B_{(s)}\bar{B}_{(s)}$ decays, and due to the limited data-set size, no clear $D_{s}^{+}$, $D^{+}$, or $D_{s}^{*+}$ signals are observed in this data sample. In the $\sqrt{s}$ = 10.580 and 10.867~GeV data samples, evident $D_{s}^{+}$ and $D^{+}$ signals, and weak $D_{s}^{*+}$ signals are seen. In the fits, the $D_{s}^{+}$ and $D^{+}$ signal shapes are described by double-Gaussian functions, and the $D_{s}^{*+}$ signal shape is described by a Novosibirsk function~\cite{Novosibirsk}, where the values of parameters are fixed to those obtained from fits to corresponding signal MC distributions. The backgrounds are parametrized by first-order polynomial functions for $D_{s}^{+}$ and $D^{+}$, and a second-order polynomial function for $D_{s}^{*+}$.

\begin{figure*}[htbp]
	\begin{center}
		\includegraphics[width=5.9cm]{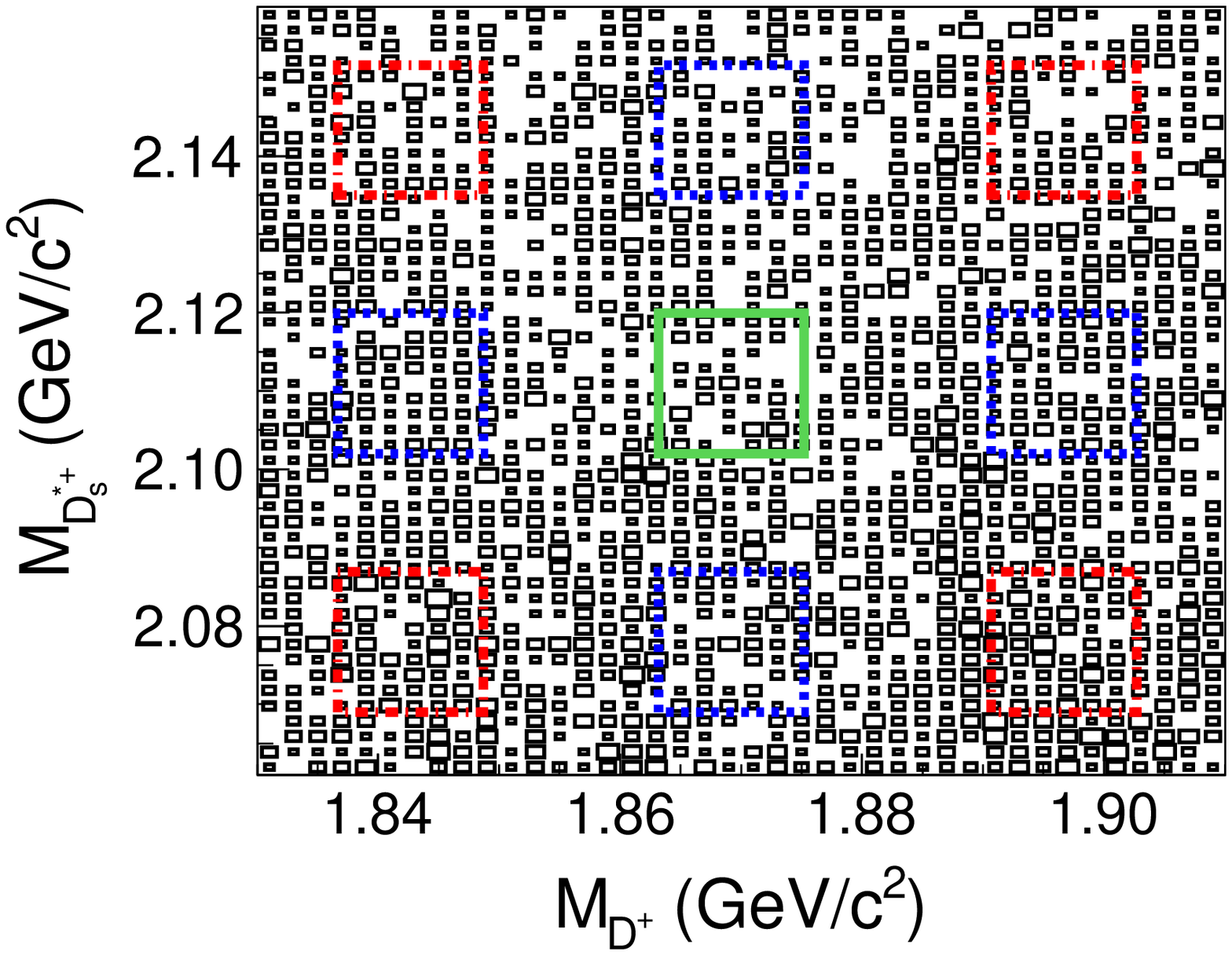}
		\includegraphics[width=5.9cm]{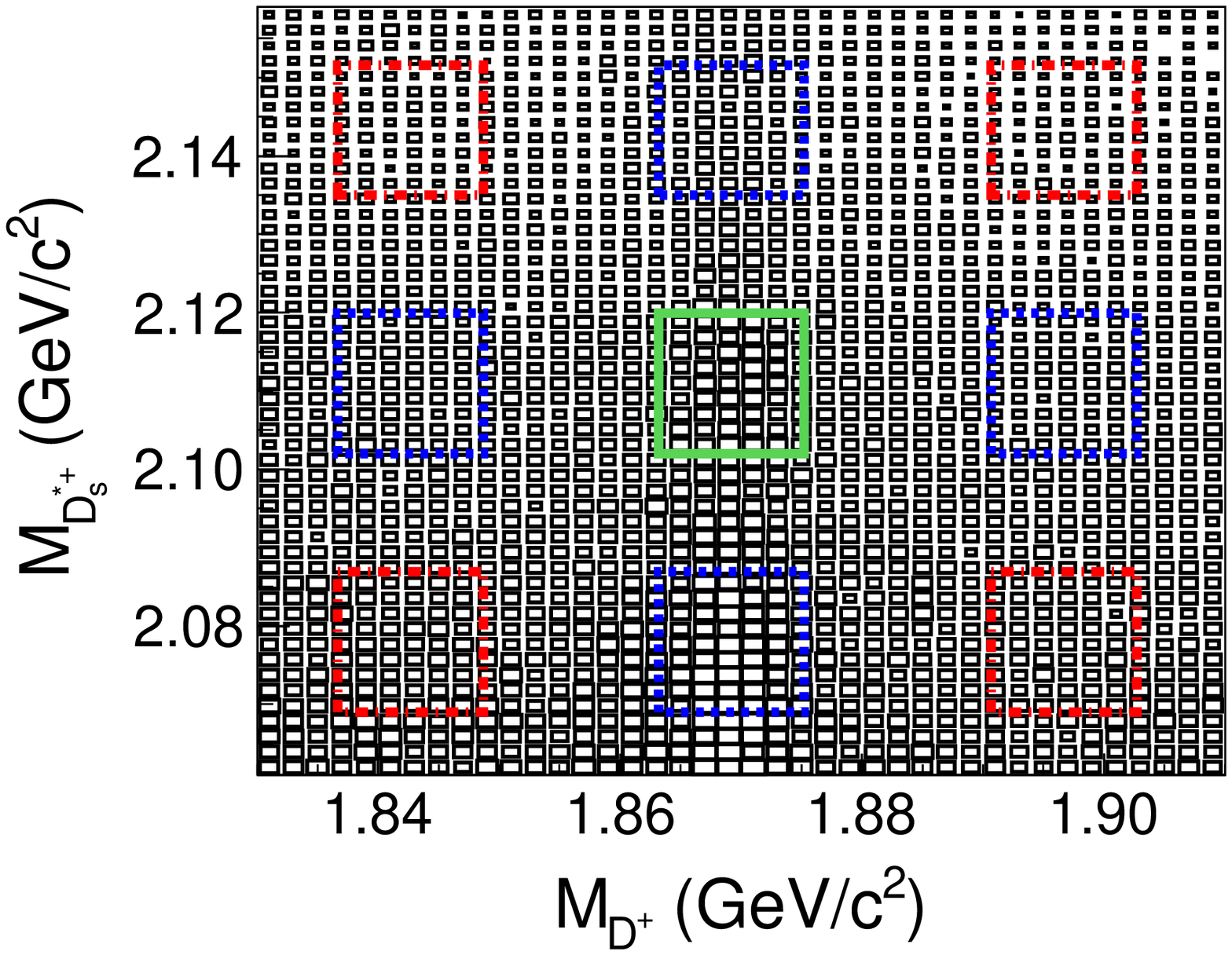}
 	   	\includegraphics[width=5.9cm]{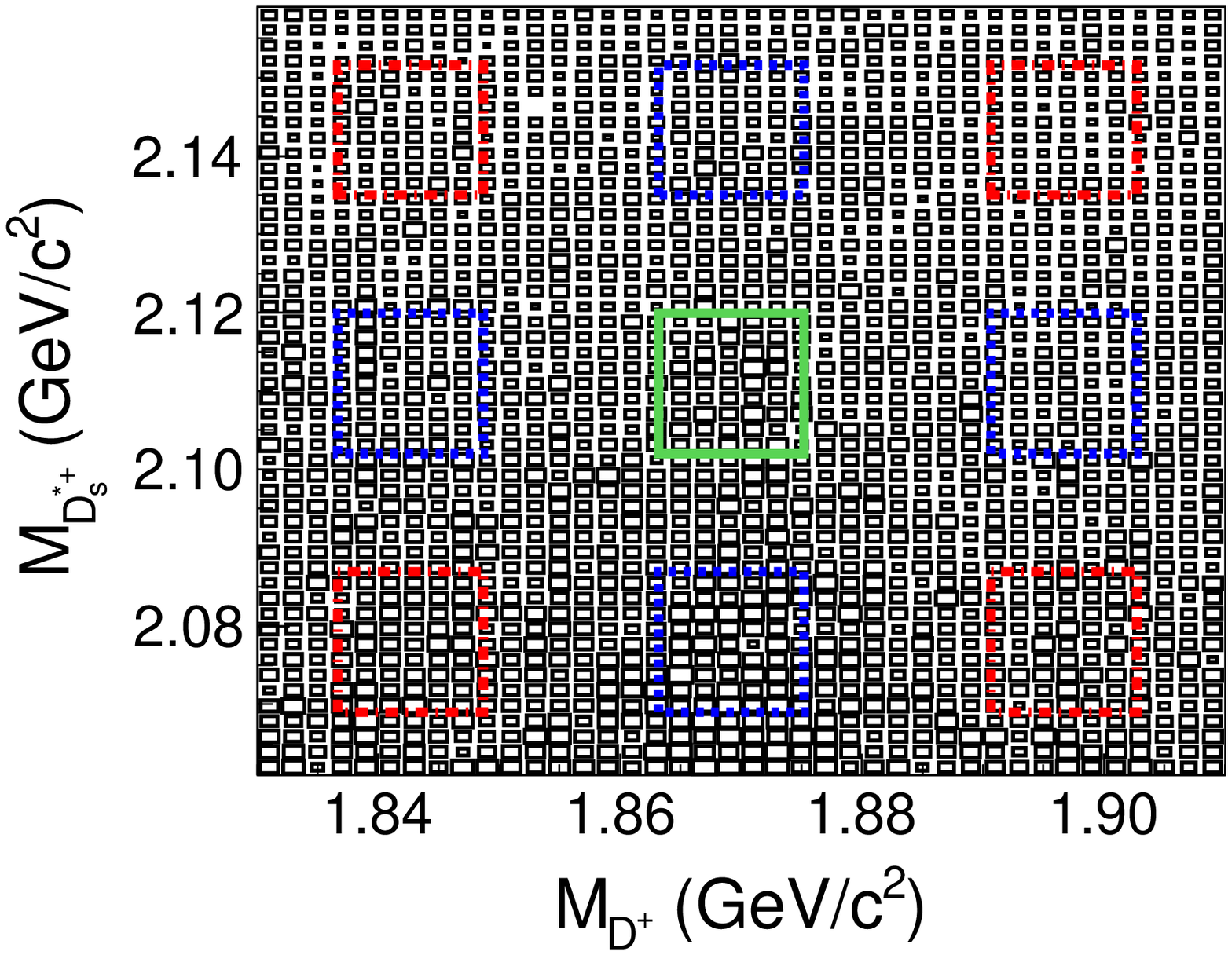}
    	\put(-495,108){\bf (a)} \put(-325,108){\bf (b)}  \put(-155,108){\bf (c)}
		\caption{The scatter plots of $M_{D_{s}^{*+}}$ versus $M_{D^{+}}$ from (a) $\sqrt{s}$ = 10.520~GeV,
			(b) $\sqrt{s}$ = 10.580~GeV, and (c) $\sqrt{s}$ = 10.867~GeV data samples.
			The central solid boxes define the signal regions, and the red dash-dotted and blue dashed boxes
			show the $M_{D^{+}}$ and $M_{D_{s}^{*+}}$ sideband regions described in the text.}\label{ee_2d}
	\end{center}
\end{figure*}

The scatter plots of $M_{D_{s}^{*+}}$ versus $M_{D^{+}}$ from the $\sqrt{s}$ = 10.520, 10.580, and 10.867~GeV data samples
are shown in Figs.~\ref{ee_2d}(a), (b), and (c), respectively. The central solid boxes show the
$D^{+}$ and $D_{s}^{*+}$ signal regions, and the blue dashed and red dash-dotted boxes show the $M_{D^{+}}$ and $M_{D_{s}^{*+}}$
sidebands. The background contribution from the normalized $M_{D^{+}}$ and $M_{D_{s}^{*+}}$ sidebands is estimated
using the same method as described in Sec. IV.

\begin{figure*}[htbp]
	\begin{center}
		\includegraphics[width=5.9cm]{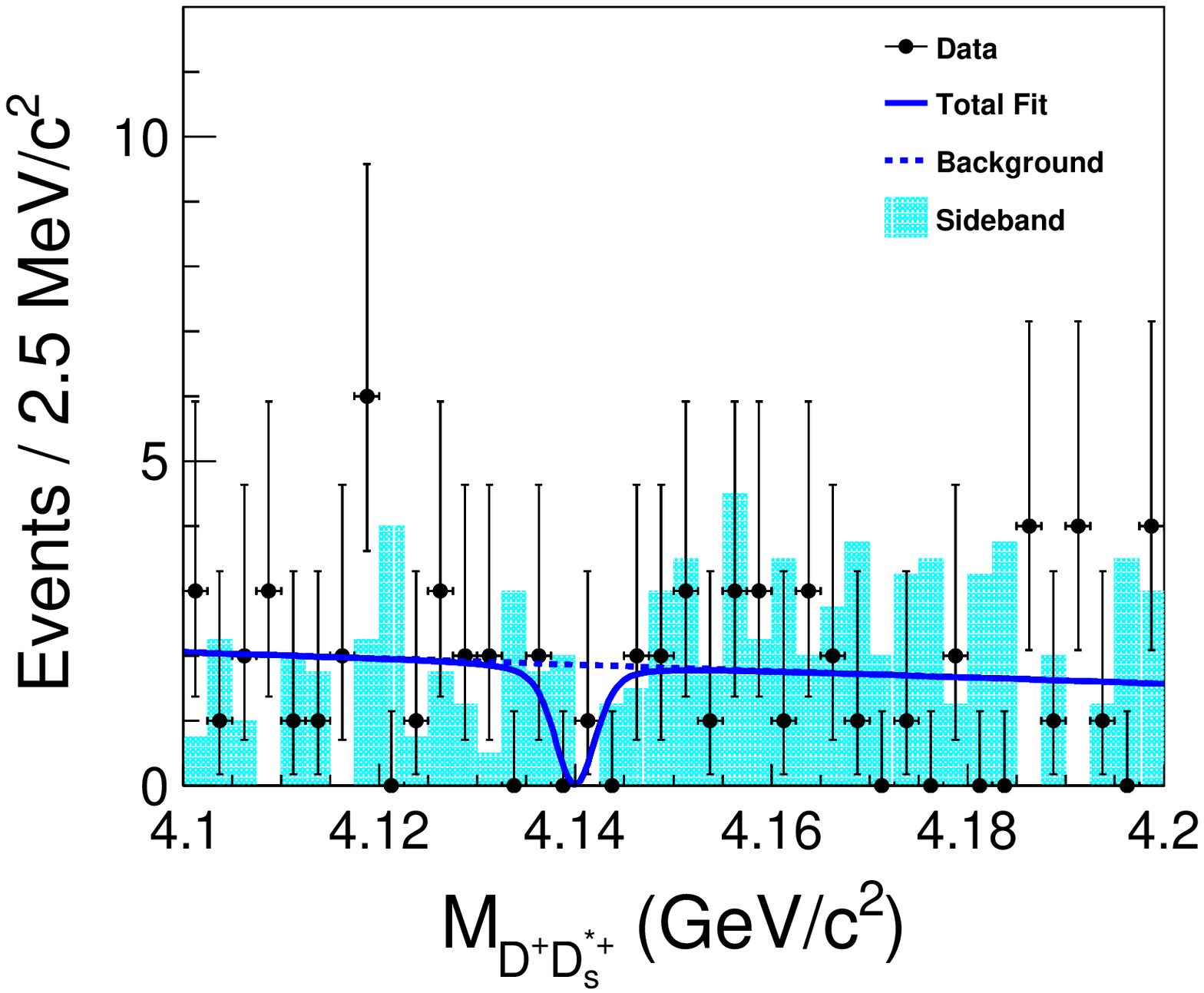}
		\includegraphics[width=5.9cm]{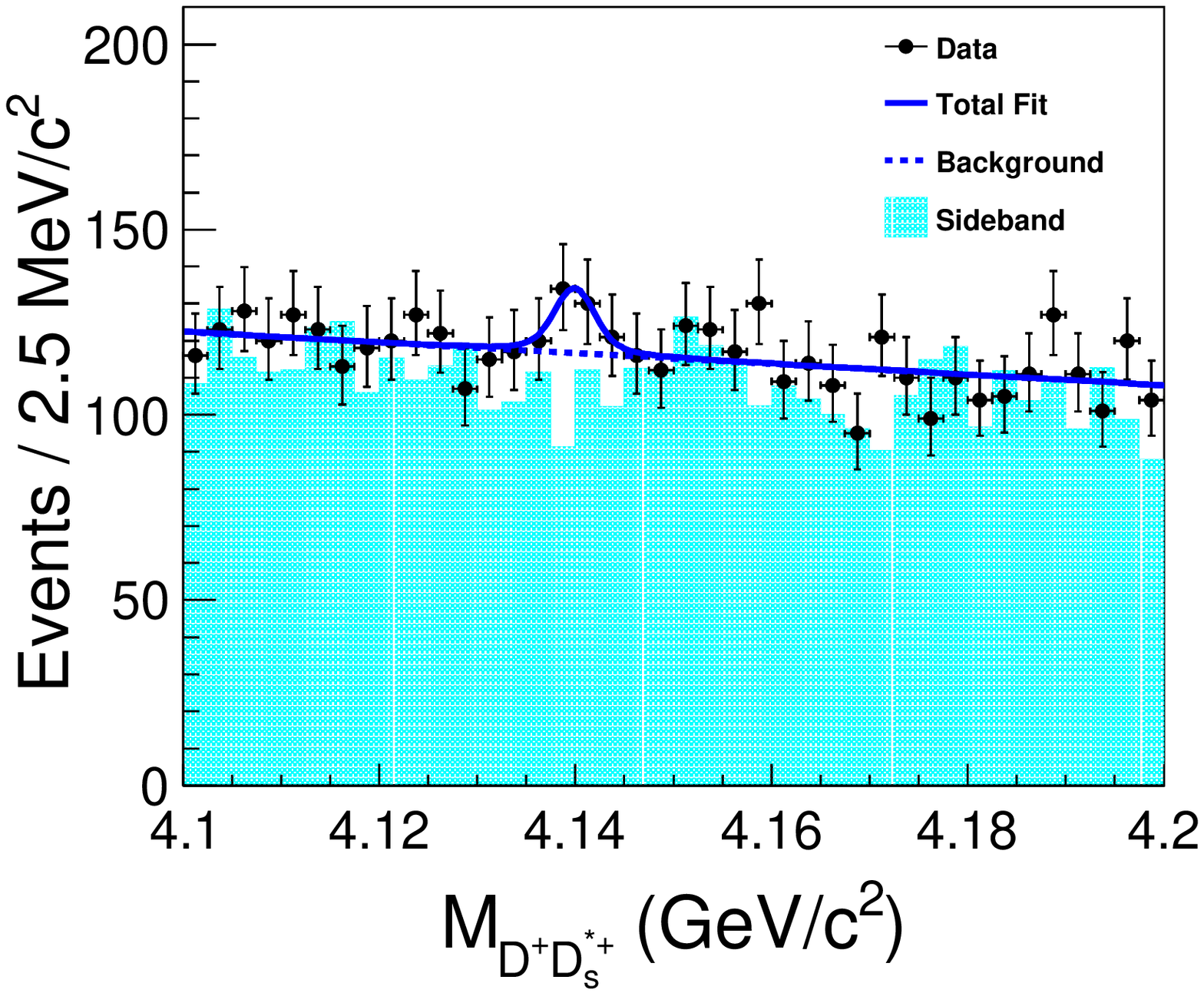}
		\includegraphics[width=5.9cm]{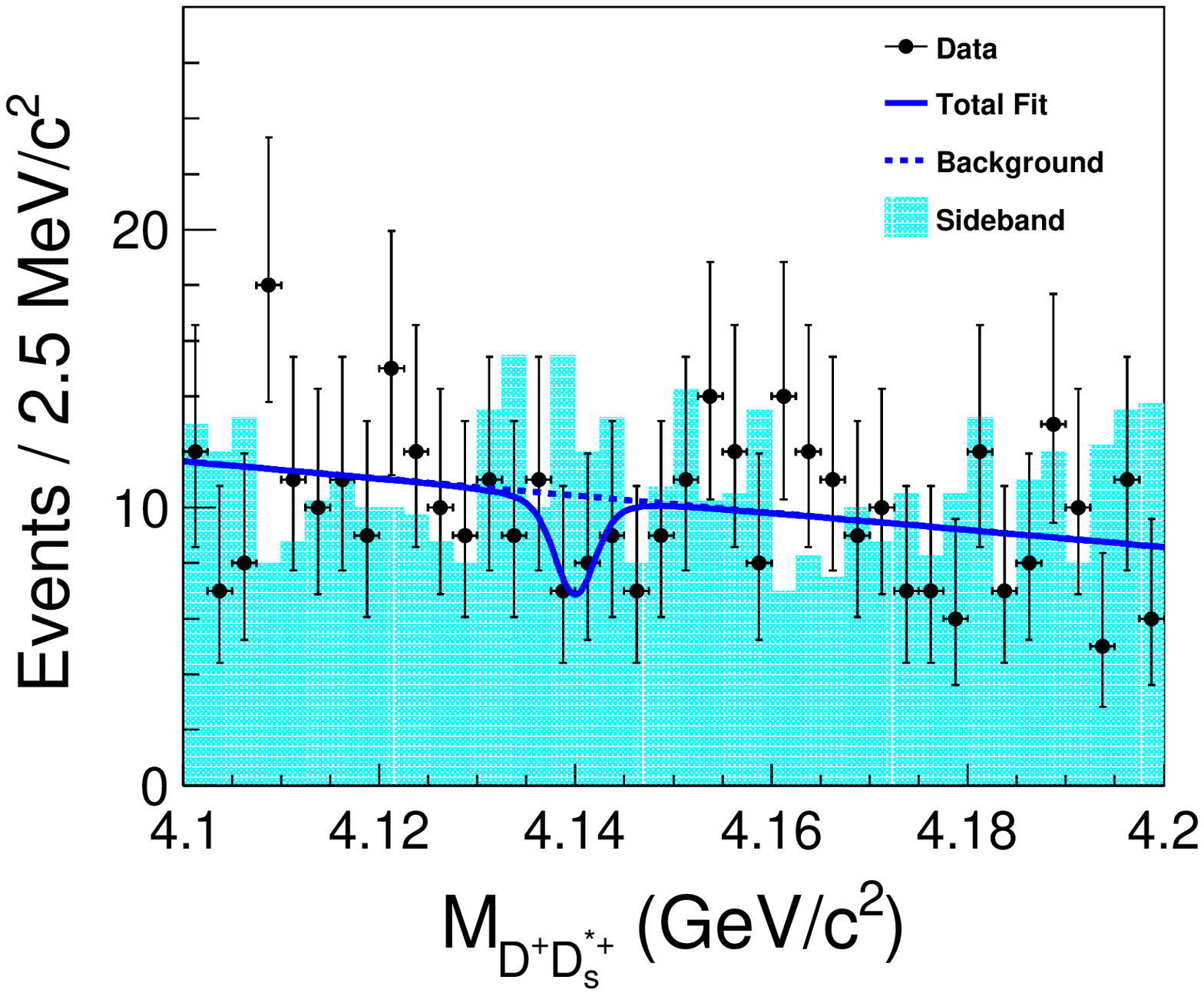}
		\put(-464,100){\bf (a)} \put(-294,100){\bf (b)}  \put(-124,100){\bf (c)}
		\caption{The invariant-mass spectra of the $D^{+}D_{s}^{*+}$ from $\EE$ annihilations at  (a) $\sqrt{s}$ = 10.520~GeV, (b) $\sqrt{s}$ = 10.580~GeV, and (c) $\sqrt{s}$ = 10.867~GeV data samples. The cyan shaded histograms are from the normalized $M_{D^+}$ and $M_{D_{s}^{*+}}$ sideband events.
		The blue solid curves show the fitted results with the $\R$ mass fixed at 4.14~GeV/$c^2$ and width fixed at 2~MeV, and the blue dashed curves are the fitted backgrounds.
		}\label{DDs_mass_ee}
	\end{center}
\end{figure*}

Figure~\ref{DDs_mass_ee} shows the invariant-mass distributions of $D^{+}D_{s}^{*+}$
from $\sqrt{s}$ = 10.520, 10.580, and 10.867~GeV data samples,
respectively, together with the backgrounds from the normalized $M_{D^{+}}$ and $M_{D_{s}^{*+}}$ sidebands.
There are no significant signals for $\R$ states in any of the data samples.
An unbinned extended maximum-likelihood fit is performed to the $M_{D^{+}D_{s}^{*+}}$ distribution in a way similar
to the methods in Sec. IV.
The fitted results with the $M_{\R}$ fixed at 4.14~GeV/$c^2$ and $\Gamma_{R^{++}}$ fixed at 2~MeV are shown in Fig.~\ref{DDs_mass_ee}
as an example. The local $\R$ significance is calculated using the same method as described in Sec. IV.
The fitted $\R$ signal yields at typically assumed mass points with $\Gamma_{\R}$ fixed at values ranging from 0 to 5~MeV in steps of 1~MeV and the corresponding statistical significances are listed in Table~\ref{tab:ee}.

The product of Born cross section and branching fraction $\sigma(\EE \to R^{++} + anything) \times \BR(R^{++} \to D^{+} D_{s}^{*+})$ is calculated from the following formula: $$\frac{N^{\rm fit} \times |1-\prod|^{2}}{\lum \times \sum_{i}\varepsilon_{i}\BR_{i}\times (1+\delta)_{\rm ISR}},$$
where $N^{\rm fit}$ is the number of fitted signal yields in data, $|1-\prod|^{2}$ is the vacuum polarization factor,
$\lum$ is the integrated luminosity, the index $i$ runs for all final-state modes with $\varepsilon_i$ being the corresponding efficiency and ${\cal B}_i$
the product of all secondary branching fractions of the mode $i$, and $(1+\delta)_{\rm ISR}$ is the radiative correction factor.
The radiative correction factors $(1+\delta)_{\rm ISR}$ are 0.710, 0.710, and 0.707 calculated using formulae given in Ref.~\cite{ISR}
for $\sqrt{s}$ = 10.520, 10.580, and 10.867~GeV, respectively; the values of $|1-\prod|^{2}$~\cite{vacuum} are 0.931, 0.930, and 0.929 for
$\sqrt{s}$ = 10.520, 10.580, and 10.867~GeV. In the calculation of $(1+\delta)_{\rm ISR}$, we assume that the
dependence of the cross section on $s$ is 1/$s$. The calculated values of $\sigma(\EE \to R^{++} + anything) \times \BR(R^{++} \to D^{+} D_{s}^{*+})$
at $\sqrt{s}$ = 10.520,  10.580, and 10.867~GeV under typical assumptions of $\R$ mass are listed in Table~\ref{tab:ee}.

\begin{figure*}[htbp]
	\begin{center}
		\includegraphics[width=5.9cm]{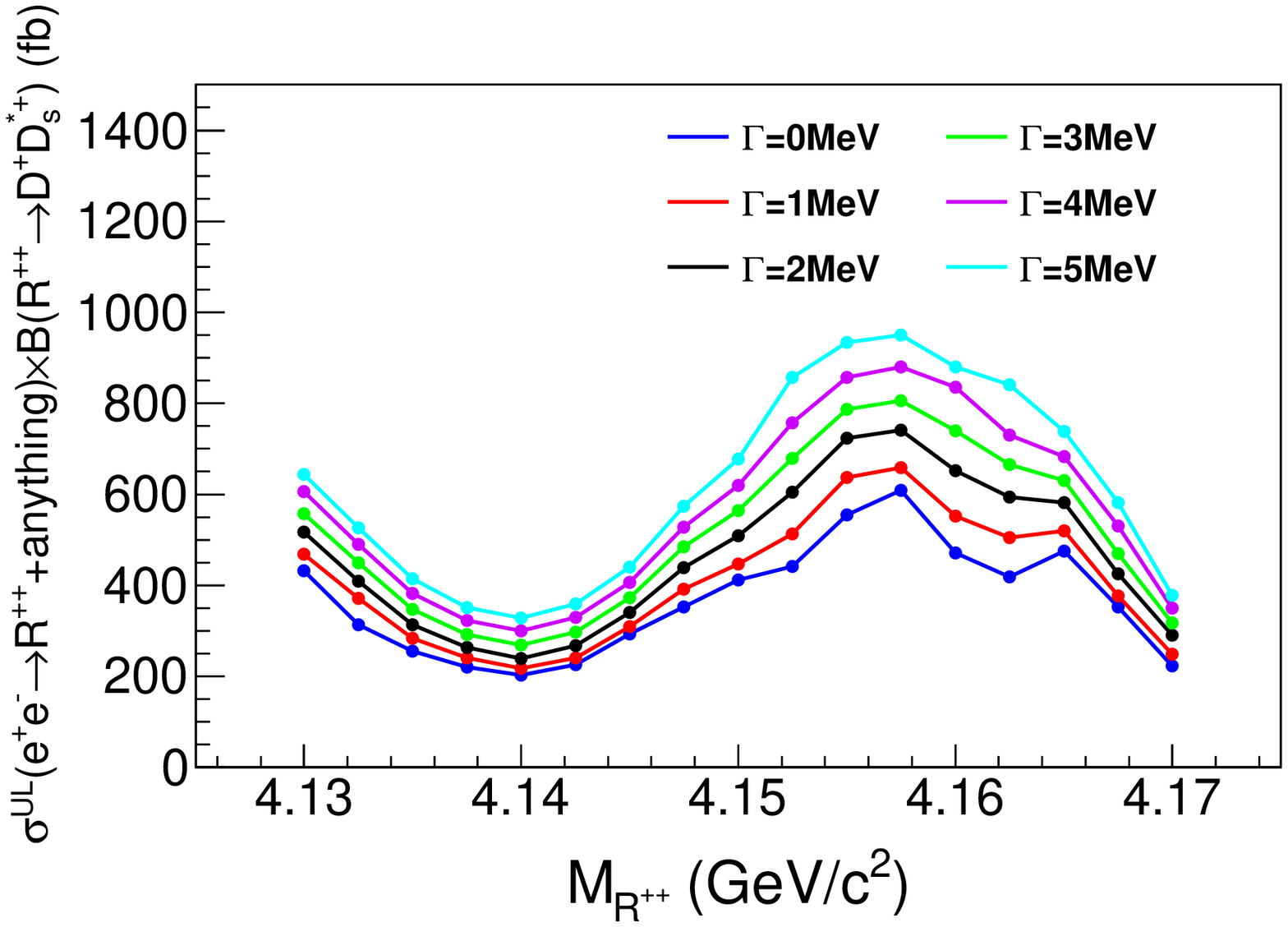}
		\includegraphics[width=5.9cm]{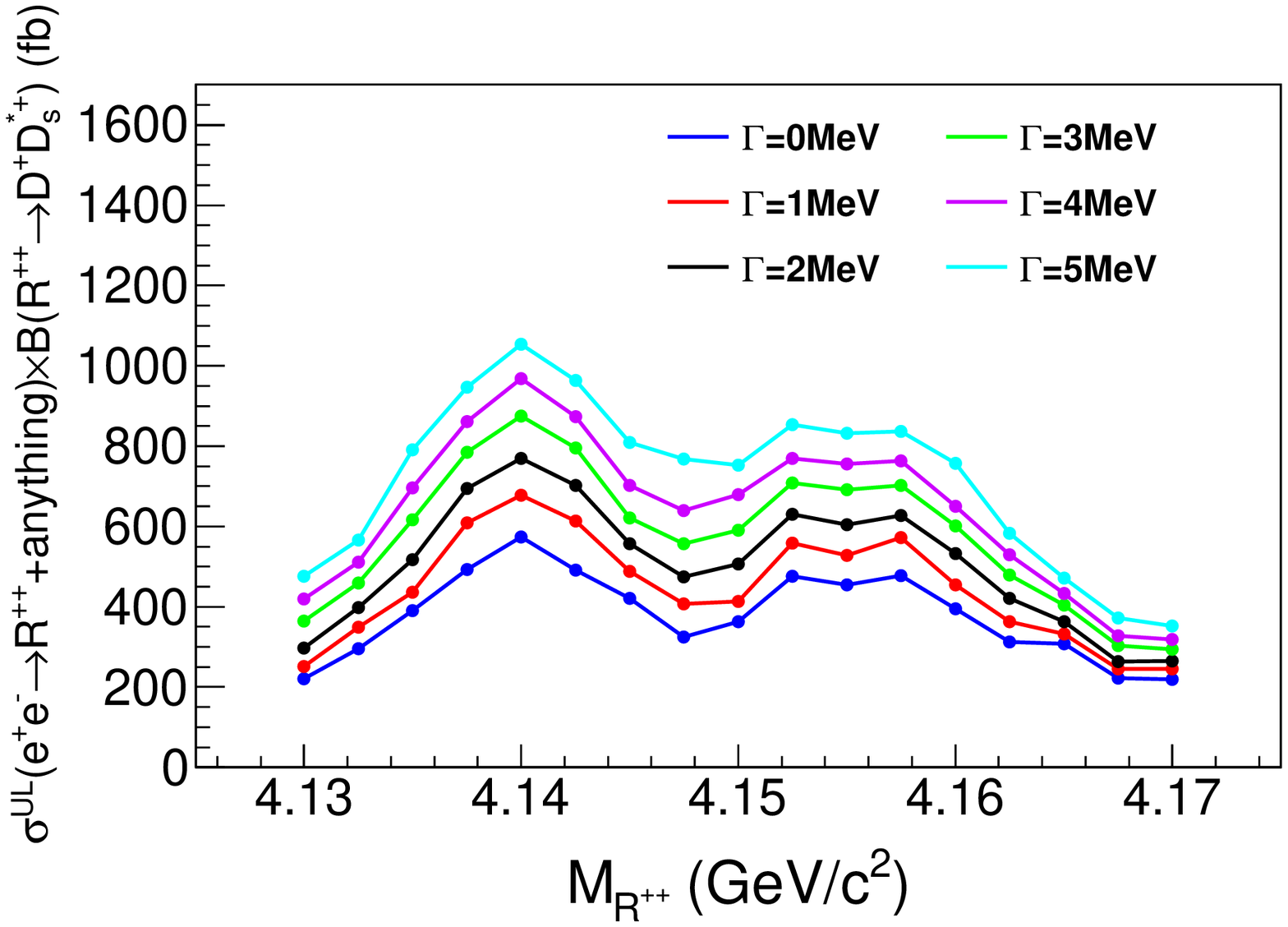}
		\includegraphics[width=5.9cm]{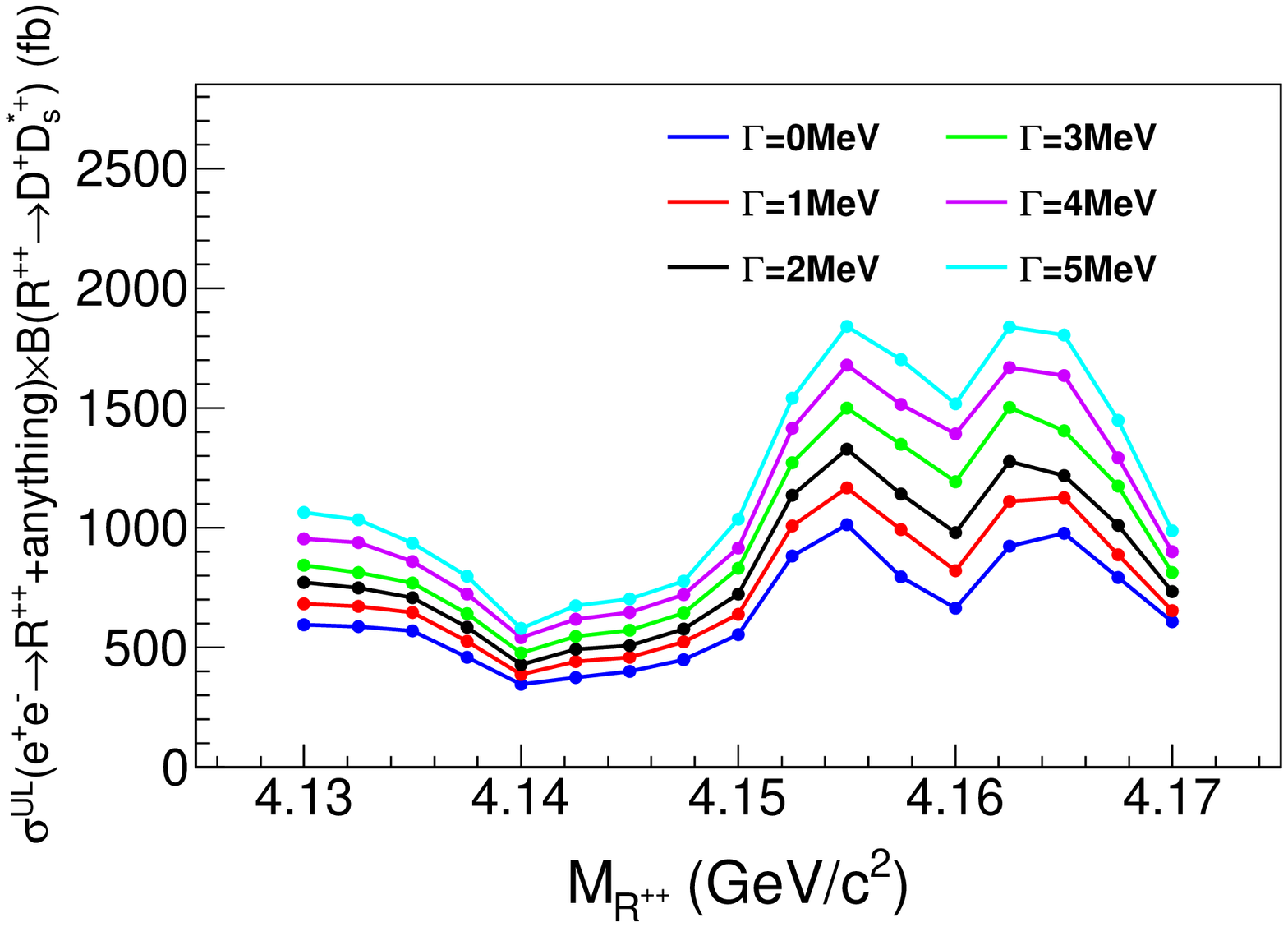}
		\put(-464,90){\bf (a)} \put(-294,90){\bf (b)}  \put(-124,90){\bf (c)}
		\caption{The 90\% C.L. upper limits on the product values of the $\EE \to R^{++} + anything$ cross sections and the branching fraction of $\R \to D^+D_s^{*+}$ at (a) $\sqrt{s}$ = 10.520~GeV, (b) $\sqrt{s}$ = 10.580~GeV, and (c) $\sqrt{s}$ = 10.867~GeV as a function of the assumed $R^{++}$ masses with widths varying from 0 to 5~MeV in steps of 1~MeV.}\label{ee_up}
	\end{center}
\end{figure*}

Since the statistical significance in each case is less than 3$\sigma$, Bayesian upper limits at the 90\% C.L.
on $N^{\rm UL}$ are obtained using the same method as described in Sec. IV.
The results for  $N^{\rm UL}$ and product values of Born cross section and branching
fraction ($\sigma^{\rm UL}(\EE \to R^{++} + anything) \times \BR(R^{++} \to D^{+} D_{s}^{*+})$) in $\EE$ collisions at $\sqrt{s}$ = 10.520, 10.580,
and 10.867~GeV under typical assumptions of $\R$ mass with $\Gamma_{R^{++}}$ fixed at values ranging from 0 to 5~MeV are listed in
Table~\ref{tab:ee}. The 90\% C.L. upper limits on the product values of the
$\EE \to \R + anything$ cross sections and the branching fraction of $\R \to D^{+}D_{s}^{*+}$ at $\sqrt{s}$ = 10.520,
10.580, and 10.867~GeV for all hypothetical $\R$ masses with widths varying from 0 to 5~MeV are shown in
Figs.~\ref{ee_up}(a)$-$(c), respectively.

\begin{sidewaystable*}[htbp]
	\vspace{8cm}
	\caption{\label{tab:ee} Summary of the 90\% C.L. upper limits on the product values of Born cross sections and the branching fractions for $\EE \to \R + anything$ at $\sqrt{s}$ = 10.520, 10.580, and 10.867~GeV with $\R \to D^{+}D_{s}^{*+}$ under typical assumptions of $\R$ mass ($M_{\R}$ in GeV/$c^2$) and width ($\Gamma_{R^{++}}$ in MeV) as examples, where $N^{\rm fit}$ is the number of  fitted signal events, $N^{\rm UL}$ is the 90\% C.L. upper limit on the number of signal events taking into account systematic uncertainties, $\Sigma(\sigma)$ is the local $\R$ significance, $\Sigma_{i}(\epsilon_i\BR_i)$ is the sum of
	product of the detection efficiency and the product of all secondary branching fractions for each reconstruction mode, $\sigma_{\rm multi}$ is the total multiplicative systematic uncertainty, $\sigma_{\rm add}$ is the additive systematic uncertainty, $\sigma \times \BR$ ($\sigma(\EE \to R^{++} + anything) \times \BR(R^{++} \to D^{+} D_{s}^{*+})$) is the product value of Born cross section and branching fraction, and $\sigma^{\rm UL} \times \BR$ ($\sigma^{\rm UL}(\EE \to R^{++} + anything) \times \BR(R^{++} \to D^{+} D_{s}^{*+})$) is the 90\% C.L. upper limit on the  product value of Born cross section and  branching fraction with systematic uncertainties included.}
	\linespread{1.2}
	\scriptsize
	\vspace{0.5cm}
     \begin{tabular}{ccc@{$/$}c@{$/$}cc@{$/$}c@{$/$}cc@{$/$}c@{$/$}cccc@{$/$}c@{$/$}cc@{$/$}c@{$/$}cc@{$/$}c@{$/$}c}
	\hline\hline
	\multicolumn{22}{c}{\scriptsize{$\EE \to \R + anything$ at $\sqrt{s}$ = 10.520/10.580/10.867 GeV, $\R \to D^{+}D_{s}^{*+}$}} \\\hline
    $M_{\R}$ & $\Gamma_{R^{++}}$ & \multicolumn{3}{c}{ $N^{\rm fit}$} &  \multicolumn{3}{c}{$N^{\rm UL}$} & \multicolumn{3}{c}{$\Sigma(\sigma)$} & $\Sigma_{i}(\epsilon_i\BR_i)$ ($\times10^{-5}$) & $\sigma_{\rm multi}(\%)$ & \multicolumn{3}{c}{$\sigma_{\rm add}(\%)$} & \multicolumn{3}{c}{$\sigma \times \BR$ (fb)} & \multicolumn{3}{c}{ $\sigma^{\rm UL} \times \BR$ (fb)} \\
	\hline
4.13& 0 &1.4$\pm$2.3 & $-24.6\pm$17.8 &  0.4$\pm$5.2 & 6.6& 22.6& 11.2 & 0.7 & -   &0.1 & 22.4/18.9/20.4 &7.9&  6.7&8.7 &6.7 &  91.6$\pm$150.4& $-239.8\pm$173.5 &21.2$\pm$275.9  & 431.7 &220.3 &594.2 \\
4.13& 1 &1.3$\pm$2.5 & $-25.5\pm$19.9 &  0.4$\pm$5.9 & 7.1& 25.6& 12.4 & 0.5 & -   &0.1 & 22.2/18.8/19.7 &7.9&  8.3&9.2 &6.3 &  85.8$\pm$165.0& $-249.9\pm$195.0 &22.0$\pm$324.2  & 468.6 &250.9 &681.3 \\
4.13& 2 &1.0$\pm$2.9 & $-27.0\pm$22.9 &  0.1$\pm$6.7 & 7.8& 30.1& 14.2 & 0.4 & -   &0.1 & 22.1/18.7/19.9 &7.9& 11.8&9.9 &6.5 &  66.3$\pm$192.3& $-266.0\pm$225.6 &5.4$\pm$364.4   & 517.1 &296.5 &772.3 \\
4.13& 3 &0.8$\pm$3.1 & $-27.7\pm$26.8 & $-0.2\pm$7.7 & 8.3& 36.6& 15.6 & 0.3 & -   & -  & 21.8/18.5/20.0 &7.9& 12.9&11.2&6.9 &  53.8$\pm$208.3& $-275.8\pm$266.9 &$-10.8\pm$416.7 & 557.8 &364.5 &844.3 \\
4.13& 4 &0.7$\pm$3.4 & $-27.5\pm$30.0 & $-0.5\pm$8.5 & 8.9& 42.6& 17.1 & 0.2 & -   & -  & 21.5/18.7/19.4 &7.9& 14.2&13.5&7.3 &  47.7$\pm$231.7& $-270.9\pm$295.6 &$-27.9\pm$474.2 & 606.5 &419.7 &954.1 \\
4.13& 5 &0.3$\pm$3.7 & $-27.5\pm$32.8 & $-0.7\pm$9.5 & 9.4& 47.8& 18.8 & 0.1 & -   & -  & 21.4/18.5/19.1 &7.9& 15.7&13.8&7.9 &  20.5$\pm$253.3& $-273.9\pm$326.6 &$-39.7\pm$538.4 & 643.6 &476.0 &1065.4\\
4.14& 0 &$-3.5\pm$1.2&  30.8$\pm$18.3 & $-4.7\pm$4.1 & 3.1& 58.6&  6.5 &  -  &1.7  & -  & 22.4/18.8/20.3 &7.9& 10.5&8.7 &15.9&$-228.9\pm$78.5 &  301.8$\pm$179.3 &$-250.6\pm$218.6& 202.8 &574.2 &346.6 \\
4.14& 1 &$-4.0\pm$1.3&  37.4$\pm$21.4 & $-7.1\pm$4.8 & 3.3& 68.8&  7.1 &  -  &1.8  & -  & 22.2/18.7/19.8 &7.9& 11.8&6.8 &10.2&$-264.0\pm$85.8 &  368.5$\pm$210.8 &$-388.1\pm$262.4& 217.8 &677.8 &388.1 \\
4.14& 2 &$-4.6\pm$1.5&  43.3$\pm$24.2 & $-8.9\pm$7.0 & 3.6& 78.1&  7.9 &  -  &1.8  & -  & 22.0/18.7/19.9 &7.9& 12.2&7.6 &8.9 &$-306.3\pm$99.9 &  426.6$\pm$238.4 &$-484.1\pm$380.7& 239.7 &769.4 &429.7 \\
4.14& 3 &$-5.2\pm$1.7&  49.0$\pm$27.2 &$-10.9\pm$6.5 & 4.0& 87.4&  8.8 &  -  &1.8  & -  & 21.8/18.4/19.9 &7.9& 13.5&9.2 &7.9 &$-349.5\pm$114.3&  490.6$\pm$272.3 &$-592.9\pm$353.5& 268.8 &875.1 &478.6 \\
4.14& 4 &$-6.1\pm$1.9&  54.6$\pm$30.4 &$-12.5\pm$7.4 & 4.4& 97.2&  9.7 &  -  &1.8  & -  & 21.5/18.5/19.4 &7.9& 12.9&10.5&6.5 &$-415.7\pm$129.5&  543.7$\pm$302.7 &$-697.4\pm$412.9& 299.8 &967.9 &541.2 \\
4.14& 5 &$-6.6\pm$2.2&  58.7$\pm$32.9 &$-13.2\pm$7.9 & 4.8&104.7& 10.9 &  -  &1.8  & -  & 21.4/18.3/19.2 &7.9& 13.2&11.8&6.1 &$-451.9\pm$150.6&  590.9$\pm$331.2 &$-744.1\pm$445.4& 328.6 &1054.0&614.5 \\
4.15& 0 &0.2$\pm$2.3 &   2.2$\pm$17.5 &$-0.9\pm$5.3  & 6.3& 37.0& 10.4 & 0.1 &0.1  & -  & 22.4/18.8/20.3 &7.9&  6.7&10.0&7.4 &  13.1$\pm$150.4&   21.6$\pm$171.5 &$-48.0\pm$282.6 & 412.1 &362.6 &554.5 \\
4.15& 1 &0.4$\pm$2.6 &   1.8$\pm$19.9 &$-0.4\pm$6.0  & 6.8& 41.7& 11.7 & 0.1 &0.1  & -  & 22.3/18.6/19.8 &7.9&  7.5&11.4&6.8 &  26.3$\pm$170.8&   17.8$\pm$197.1 &$-21.9\pm$328.0 & 446.8 &413.0 &639.6 \\
4.15& 2 &0.5$\pm$2.9 &   4.3$\pm$23.3 & 0.1$\pm$6.9  & 7.6& 51.1& 13.3 & 0.2 &0.1  &0.1 & 21.9/18.6/19.9 &7.9&  9.3&12.3&5.3 &  33.4$\pm$194.0&   42.6$\pm$230.8 &5.4$\pm$375.3   & 508.4 &506.1 &723.4 \\
4.15& 3 &0.6$\pm$3.3 &   7.0$\pm$26.3 & 0.7$\pm$7.8  & 8.4& 59.0& 15.2 & 0.2 &0.1  &0.1 & 21.8/18.4/19.8 &7.9& 11.9&13.5&5.1 &  40.3$\pm$221.8&   70.1$\pm$263.3 &38.3$\pm$426.4  & 564.5 &590.7 &830.9 \\
4.15& 4 &0.6$\pm$3.5 &  11.3$\pm$29.4 & 0.9$\pm$8.5  & 9.1& 67.8& 16.5 & 0.2 &0.5  &0.1 & 21.5/18.4/19.5 &7.9& 13.4&13.8&4.8 &  40.9$\pm$238.5&  113.8$\pm$294.4 &50.0$\pm$471.8  & 620.1 &678.8 &915.9 \\
4.15& 5 &0.7$\pm$3.9 &  15.4$\pm$32.0 & 1.7$\pm$9.4  & 9.9& 74.0& 18.5 & 0.2 &0.5  &0.2 & 21.4/18.1/19.3 &7.9& 15.3&14.7&5.2 &  47.9$\pm$267.0&  156.7$\pm$325.7 &95.3$\pm$527.2  & 677.8 &753.2 &1037.5\\
4.16& 0 &0.4$\pm$2.6 &   9.1$\pm$17.9 & 2.5$\pm$5.5  & 7.2& 40.1& 12.4 & 0.2 &0.5  &0.5 & 22.4/18.7/20.2 &7.9& 10.2&14.8&11.6&  26.2$\pm$170.1&   89.7$\pm$176.3 &134.0$\pm$294.7 & 470.9 &395.1 &664.4 \\
4.16& 1 &1.1$\pm$3.1 &  10.3$\pm$20.3 & 4.3$\pm$6.4  & 8.4& 45.6& 15.1 & 0.4 &0.5  &0.7 & 22.3/18.5/19.9 &7.9&  9.7&13.7&10.5&  72.3$\pm$203.7&  102.6$\pm$202.2 &233.9$\pm$348.1 & 551.9 &454.1 &821.3 \\
4.16& 2 &1.9$\pm$3.5 &  12.7$\pm$23.9 & 6.2$\pm$7.5  & 9.7& 53.8& 18.0 & 0.6 &0.5  &0.9 & 21.8/18.6/19.9 &7.9& 11.4&14.5&9.7 &  127.7$\pm$235.2& 125.8$\pm$236.7 &337.2$\pm$407.9 & 651.9 &532.9 &979.0 \\
4.16& 3 &2.7$\pm$3.9 &  14.3$\pm$26.8 & 9.1$\pm$8.6  &11.0& 59.8& 21.8 & 0.7 &0.5  &1.1 & 21.8/18.3/19.8 &7.9& 12.5&15.3&10.4&  181.5$\pm$262.1& 144.0$\pm$269.8 &497.5$\pm$470.1 & 739.3 &602.0 &1191.7\\
4.16& 4 &3.3$\pm$4.3 &  14.3$\pm$29.0 &11.5$\pm$9.6  &12.2& 64.2& 25.1 & 0.8 &0.5  &1.3 & 21.4/18.2/19.5 &7.9& 13.2&15.9&12.3&  225.9$\pm$294.4& 144.8$\pm$293.5 &638.3$\pm$532.9 & 835.2 &649.9 &1393.2\\
4.16& 5 &3.5$\pm$4.5 &  18.1$\pm$32.5 &13.1$\pm$10.3 &12.8& 73.5& 28.5 & 0.8 &0.6  &1.3 & 21.3/17.9/19.4 &7.9& 13.1&16.3&13.2&  240.7$\pm$209.5& 186.3$\pm$334.5 &730.9$\pm$574.7 & 880.4 &756.5 &1590.1\\
4.17& 0 &$-2.2\pm$1.3& $-16.0\pm$17.2 & 1.7$\pm$5.3  & 3.4& 22.1& 11.3 &  -  & -   &0.3 & 22.4/18.6/20.1 &7.9&  8.5&11.9&10.8& $-143.9\pm$85.0 &$-158.5\pm$170.4 &91.5$\pm$285.4  & 222.4 &218.9 &608.5 \\
4.17& 1 &$-2.7\pm$1.5& $-20.9\pm$20.0 & 1.4$\pm$5.9  & 3.8& 24.5& 12.1 &  -  & -   &0.2 & 22.4/18.4/20.0 &7.9&  6.7&10.8&12.7& $-176.6\pm$98.1 &$-209.3\pm$200.2 &75.8$\pm$319.3  & 248.5 &245.3 &654.8 \\
4.17& 2 &$-3.3\pm$1.8& $-27.1\pm$22.9 & 1.3$\pm$6.7  & 4.3& 26.8& 13.5 &  -  & -   &0.2 & 21.7/18.6/19.9 &7.9&  5.8&12.4&13.5& $-222.8\pm$121.5&$-268.4\pm$226.8 &70.7$\pm$364.4  & 290.3 &265.4 &734.3 \\
4.17& 3 &$-3.8\pm$2.1& $-33.1\pm$25.9 & 1.1$\pm$7.5  & 4.7& 29.1& 14.8 &  -  & -   &0.1 & 21.7/18.2/19.7 &7.9&  4.1&13.2&14.9& $-256.6\pm$141.8&$-335.1\pm$262.2 &60.4$\pm$412.1  & 317.3 &294.6 &813.2 \\
4.17& 4 &$-4.2\pm$2.5& $-37.3\pm$28.3 & 1.0$\pm$8.5  & 5.1& 31.1& 16.2 &  -  & -   &0.1 & 21.4/18.0/19.5 &7.9&  3.3&13.9&16.1& $-287.5\pm$171.2&$-381.8\pm$289.6 &55.5$\pm$471.8  & 349.2 &318.3 &899.2 \\
4.17& 5 &$-4.6\pm$2.8& $-43.5\pm$31.6 & 1.1$\pm$9.3  & 5.5& 33.8& 17.8 & -   & -   &0.1 & 21.3/17.7/19.5 &7.9&  2.3&15.8&16.4& $-316.4\pm$192.6&$-452.8\pm$328.9 &61.1$\pm$516.2  & 378.3 &351.8 &988.0 \\
\hline\hline
\end{tabular}
\end{sidewaystable*}

\section{\boldmath Systematic Uncertainties}
The systematic uncertainties in the branching fraction and Born
cross section measurements can be divided into two categories:
multiplicative systematic uncertainties and additive systematic uncertainties.
	
The sources of multiplicative systematic uncertainties include detection-efficiency-related uncertainties,
the statistical uncertainty of the MC efficiency, the modeling of MC event generation,
branching fractions of intermediate states, energy dependence of the cross sections,
the total numbers of $\onetwos$ events as well as the integrated luminosity.

The detection-efficiency-related uncertainties include those for
tracking efficiency (0.35\% per track), particle identification efficiency
(1.8\% per kaon, 1.0\% per pion), as well as  momentum-weighted $K_{S}^{0}$
selection efficiency (2.2\%)~\cite{kserr}. The photon reconstruction contributes
2.0\% per photon, as determined from radiative Bhabha events. The above individual
uncertainties from different reconstructed modes are added linearly,
weighted by the product of the detection efficiency and the product of all secondary branching fractions ($\epsilon_{i}\times\BR_{i}$).
Assuming these uncertainties are independent and adding them in quadrature,  the final uncertainty related to the reconstruction efficiency
is 6.6\%.

The MC statistical uncertainties are estimated using the yields of selected and generated events;
these are 1.0\% or less. We use the {\sc EvtGen} generator to generate the signal
MC samples. By changing the recoil mass of the $\R$, the efficiencies are changed
by $(1-3)$\%. To be conservative, we take 1\% and 3\% as the systematical uncertainties
related to signal MC statistics and generation.

The relative uncertainties of branching fractions for
$D^+ \to K^{-} \pip \pip$, $D^+ \to K_{S}^{0} \pip$,  $K_{S}^{0}\to \pip\pim$, $D_{s}^{*+} \to \gamma D_{s}^{+}$, $D_{s}^+ \to \phi(\to K^{+}K^{-})\pip$,
and $D_{s}^+ \to \bar{K}^{*}(892)^{0}(\to K^{-}\pip)K^{+}$ are taken from Ref.~\cite{PDG} and summed in quadrature
to obtain the total uncertainty of the branching fractions of the intermediate states for each reconstructed mode.
The above individual uncertainties from different reconstructed modes are added linearly with a weighting factor
of $\epsilon_{i}\times\BR_{i}$ to obtain 2.5\% as the uncertainty due to the branching fractions of
intermediate states.

Changing the $s$ dependence of the cross sections of $\EE \to \R + anything$ from 1/$s$ to 1/$s^4$, the radiative correction
factors $(1+\delta)_{\rm ISR}$  become 0.712, 0.711, and 0.709 for $\sqrt{s}$ = 10.520, 10.580, and 10.867~GeV, respectively.
The differences are less than 0.3\%. Thus, the systematic uncertainty related to the radiative correction factors
is negligible with respect to the other sources.

The uncertainties on the total numbers of $\ones$ and $\twos$  events are 2.0\% and 2.3\%, respectively, which
are mainly due to imperfect simulations of the charged track multiplicity distributions
from inclusive hadronic MC events. The total luminosity is determined to 1.4\%
precision using wide-angle Bhabha scattering events.

Additive systematic uncertainties due to the mass resolution and fit are considered as follows. The uncertainty due to the
mass resolution is studied by using the control sample of $B^0 \to D^- D_s^{*+}$; the difference
in mass resolution between MC simulation and data is around 10\%. Thus, the uncertainty due to the mass
resolution is estimated by enlarging the mass resolution by 10\% when fitting the $D^+ D_s^{*+}$ invariant-mass
distributions. To estimate the uncertainties associated with the fit, the order of the background polynomial is changed
from first to second or third and the range of the fit is changed by $\pm$30~MeV/$c^2$.

The upper limits on the branching fraction and Born cross section at the 90\% C.L. are determined and the systematic
uncertainties are taken into account in two steps. First, when we study the additive systematic uncertainties
described above, we take the most conservative upper limit at the 90\% C.L. on the number of $\R$ signal yields.
The differences between the most conservative upper limits and the nominal fits are
in the range of 2.3\% $-$ 16.4\% (see Tables~\ref{tab:1s2s} and~\ref{tab:ee} for detailed vaules), depending on the center-of-mass energy, the mass and width of the $\R$ state. Then, to take into account the multiplicative systematic uncertainties, the likelihood with the most conservative upper
limit is convolved with a Gaussian function whose width is the corresponding total multiplicative systematic uncertainty.

The sources of uncertainties are assumed independent, and the
total multiplicative systematic uncertainties are obtained by adding all uncertainties in quadrature.
The total multiplicative systematic uncertainties are listed in Tables~\ref{tab:1s2s} and~\ref{tab:ee}
for the measurements of $\BR(\onetwos \to R^{++} + anything) \times \BR(R^{++} \to D^{+} D_{s}^{*+}))$
and $\sigma(\EE \to R^{++} + anything) \times \BR(R^{++} \to D^{+} D_{s}^{*+})$ at $\sqrt{s}$ = 10.520, 10.580, and 10.867~GeV, respectively.

\section{\boldmath conclusion}
In summary, using the data samples of 102 million $\ones$ events, and 158 million $\twos$ events,
as well as 89.45 fb$^{-1}$, 711 fb$^{-1}$, and 121.06 fb$^{-1}$ collected at
$\sqrt{s}$ = 10.520, 10.580, and 10.867~GeV, we search for the doubly-charged $DDK$
bound state decaying to  $D^{+} D_{s}^{*+}$, referred to as $\R$, both in $\onetwos$ inclusive decays and in $\EE$ annihilations.
No evident signals are observed in all studied reactions.
We determine the 90\% C.L.\ upper limits on $\BR(\onetwos \to R^{++} + anything) \times \BR(R^{++} \to D^{+} D_{s}^{*+})$ and $\sigma(\EE \to R^{++} + anything) \times \BR(R^{++} \to D^{+} D_{s}^{*+})$ at $\sqrt{s}$
= 10.520, 10.580, and 10.867~GeV under different assumptions of $\R$ masses varying from 4.13 to 4.17 GeV/$c^2$ in steps of 2.5~MeV/$c^2$ and widths varying from 0 to 5~MeV in steps of 1~MeV.

\section{\boldmath ACKNOWLEDGMENTS}
We thank Professor Li-sheng Geng for useful discussions
and comments. We thank the KEKB group for the excellent operation of the
accelerator; the KEK cryogenics group for the efficient
operation of the solenoid; and the KEK computer group, and the Pacific Northwest National
Laboratory (PNNL) Environmental Molecular Sciences Laboratory (EMSL)
computing group for strong computing support; and the National
Institute of Informatics, and Science Information NETwork 5 (SINET5) for
valuable network support.  We acknowledge support from
the Ministry of Education, Culture, Sports, Science, and
Technology (MEXT) of Japan, the Japan Society for the
Promotion of Science (JSPS), and the Tau-Lepton Physics
Research Center of Nagoya University;
the Australian Research Council including grants
DP180102629, 
DP170102389, 
DP170102204, 
DP150103061, 
FT130100303; 
Austrian Science Fund (FWF);
the National Natural Science Foundation of China under Contracts
No.~11435013,  
No.~11475187,  
No.~11521505,  
No.~11575017,  
No.~11675166,  
No.~11705209,  
No.~11761141009,
No.~11975076,
No. 12005040; 
Key Research Program of Frontier Sciences, Chinese Academy of Sciences (CAS), Grant No.~QYZDJ-SSW-SLH011; 
the  CAS Center for Excellence in Particle Physics (CCEPP); 
the Shanghai Pujiang Program under Grant No.~18PJ1401000;  
the Ministry of Education, Youth and Sports of the Czech
Republic under Contract No.~LTT17020;
the Carl Zeiss Foundation, the Deutsche Forschungsgemeinschaft, the
Excellence Cluster Universe, and the VolkswagenStiftung;
the Department of Science and Technology of India;
the Istituto Nazionale di Fisica Nucleare of Italy;
National Research Foundation (NRF) of Korea Grant
Nos.~2016R1\-D1A1B\-01010135, 2016R1\-D1A1B\-02012900, 2018R1\-A2B\-3003643,
2018R1\-A6A1A\-06024970, 2018R1\-D1A1B\-07047294, 2019K1\-A3A7A\-09033840,
2019R1\-I1A3A\-01058933;
Radiation Science Research Institute, Foreign Large-size Research Facility Application Supporting project, the Global Science Experimental Data Hub Center of the Korea Institute of Science and Technology Information and KREONET/GLORIAD;
the Polish Ministry of Science and Higher Education and
the National Science Center;
Russian Science Foundation, Grant No. 18-12-00226;
University of Tabuk research grants S-0256-1438 and S-0280-1439 (Saudi Arabia);
the Slovenian Research Agency;
Ikerbasque, Basque Foundation for Science, Spain;
the Swiss National Science Foundation;
the Ministry of Education and the Ministry of Science and Technology of Taiwan;
and the United States Department of Energy and the National Science Foundation.


\begin{thebibliography}{**}
\bibitem{Novosibirsk} The Novosibirsk function is defined as $f(x)$ = exp[$-\frac{1}{2}$(ln$^2(1+\Lambda(x-x_0))/\tau^2+\tau^2)$$]$ with $\Lambda$ = sinh$(\tau\sqrt{\textrm{ln4}})/(\sigma\sqrt{\textrm{ln4}})$. The parameters represent the mean ($x_0$), the width ($\sigma$) and the tail asymmetry ($\tau$).
\end{thebibliography}
\end{document}